\documentclass{aastex631}
\usepackage{graphicx}	
\usepackage{amsmath}	
\usepackage{amssymb}	
\usepackage{subfigure}
\usepackage{enumitem}

\shorttitle{GRB 210731A: Asymmetric Jet Candidate}
\shortauthors{J.-D. Li et al.}
\graphicspath{{./}{figures/}}

\begin{document}

\title{Multiple rebrightenings in the optical afterglow of GRB 210731A: evidence for an asymmetric jet}

\author{Jin-Da Li}
\affiliation{Institute for Frontier in Astronomy and Astrophysics, Beijing Normal University, Beijing 102206, China}
\affiliation{School of Physics and Astronomy, Beijing Normal University, Beijing 100875, China}

\author{He Gao}
\affiliation{Institute for Frontier in Astronomy and Astrophysics, Beijing Normal University, Beijing 102206, China}
\affiliation{School of Physics and Astronomy, Beijing Normal University, Beijing 100875, China}

\correspondingauthor{He Gao}
\email{gaohe@bnu.edu.cn}

\author{Shunke Ai}
\affiliation{Department of Astronomy, School of Physics and Technology, Wuhan University, Wuhan 430072, China}
\affiliation{Niels Bohr International Academy and DARK, Niels Bohr Institute, University of Copenhagen, Blegdamsvej 17, 2100, Copenhagen, Denmark}

\author{Wei-Hua Lei}
\affiliation{Department of Astronomy, School of Physics, Huazhong University of Science and Technology, Wuhan, Hubei 430074, China}



\begin{abstract}

The broadband afterglow of Gamma-ray bursts (GRBs) is usually believed to originate from the synchrotron radiation of electrons accelerated by the external shock of relativistic jets. Therefore, the jet structure should have a significant impact on the GRB afterglow features. The latest observations indicate that the GRB jets may possess intricate structures, such as Gaussian structure, power-law structure, or jet-cocoon structure. Most recently, an abnormal afterglow of GRB 210731A has raised extensive attention, whose optical afterglow exhibites multiple rebrightening phenomena within 4 hours, posing a serious challenge to the standard afterglow model. Here we intend to interpret the characteristics of GRB 210731A afterglows within the framework of non-axisymmetric structured jets, where multiple distinct peaks in the afterglow light curve are caused by the uneven distribution of energy and velocity within the jet in the azimuth angle direction. Through Monte Carlo Markov Chain fitting, we show that a three-component asymmetric structured jet can well explain the multi-band afterglow data. The energy difference among the three components is about 1.5 orders of magnitude, with higher-energy components exhibiting slower speeds. The radiation contribution of each component has sequentially dominated the light curve of the afterglow, resulting in multiple peaks, with the highest peak occurring at the latest time. We suggest that in the future, polarization observations should be conducted on afterglows with multiple brightening signatures, which will help to effectively distinguish the structured jet model from other alternative models, such as energy injection, and ultimately help to determine the true configuration of jets.

\end{abstract}

\keywords{Gamma-ray bursts (GRBs)}


\section{Introduction} \label{sec:intro}
Multiwavelength observations of GRB afterglows were predicted before their first discoveries \citep{Meszaros1997ApJ,Costa1997Natur,Frail1997Natur,vanParadijs1997Natur}, based on a generic external forward shock model. This model, which is independent of the physical nature of the progenitor and central engine, assumes that a relativistic jet is launched and subsequently decelerated by a circumburst medium through a pair of external shocks - forward and reverse. It is likely that the reverse shock is short-lived, while the forward shock continues to plow into the medium as the jet is decelerated. The synchrotron radiation of electrons accelerated from the external forward shock powers the broadband electromagnetic radiation with a decreasing amplitude, forming the broadband afterglow of GRBs \cite[][for a review]{Gao2013NewAR}.

With a large sample of GRBs with both X-ray and optical afterglow data, \cite{wang2015} perform a systematic study to conclude that the simplest external forward shock models can account for the multiwavelength afterglow data of at least half of the GRBs. The remaining GRB afterglows may exhibit some features that deviate from the predictions of the simplest model, but it is these features that allow us to study the central engine and jet structure of GRBs using the afterglow. For example, through the appearance of flares in the afterglow, we can recognize the reactivation of the central engine \citep{Zhang2006ApJ}, and through the slow decay component, we can understand the continuous energy injection of the central engine and determine that some GRBs may have a magnetic star as their central engine \citep{Dai1998A&A,Zhang2006ApJ}. When more advanced modeling (e.g., long-lasting reverse shock, structured jets, arbitrary circumburst medium density profile) is invoked, \cite{wang2015} found that up to $>90\%$ of the afterglows could be interpreted within the framework of the external shock models.

Most recently, an abnormal afterglow of GRB 210731A has raised extensive attention \citep{deWet2023A&A}. GRB 210731A was a long-duration ($T_{90}=22.5~s$) gamma-ray burst discovered by the Burst Alert Telescope (BAT) aboard the Neil Gehrels \textit{Swift} Observatory \citep{Gehrels2004ApJ}. The multi-band observation of the afterglow used the wide-field, robotic MeerLICHT optical telescope in Sutherland showed a light curve with three similar peaks in brightness within the first four hours \citep{deWet2023A&A}. The first optical peak can be explained as the onset of afterglow, while the other two peaks are taken as subsequent re-brightenings. This multiple rebrightening phenomenon cannot be explained by a simple external shock model. One possible explanation is multiple energy injection \citep[e.g.,][]{Mangano2007A&A,Greiner2009ApJ,Swenson2013ApJ}. However, if one uses this model to interpret the data of GRB 210731A, it would require that the final energy injection is at least 1000 times the initial energy \citep{deWet2023A&A}, which means that the central engine's reactivation strength is 1000 times the initial activity strength, and there is no corresponding signature of reactivation in gamma-ray or X-ray emissions, which is puzzling. 

In addition to the energy injection, structured jet scenarios can also cause late-time afterglow deviations from the standard model. For example, Gaussian structured jets or power-law structured jets would cause the afterglow decay slope to be shallower \citep{Beniamini2020MNRAS,Gottlieb2021MNRAS}, while two-component structured jets would lead to a late-time rebrightening feature \citep{Huang2004ApJ,Peng2005ApJ}. However, these axially symmetric structured jets would have difficulties explaining the multiple similar rebrightening signatures in GRB 210731A. Most recently, \citet{Li2023MNRAS} discussed the potential characteristics of GRB afterglows within the framework of non-axisymmetric structured jets, whose results suggested that the uneven distribution of energy and velocity within the jet in the azimuth angle direction will cause multiple distinct peaks in the late afterglow light curve.

In this work, we intend to fit the multi-band afterglow data of GRB 210731A using Monte Carlo Markov Chain (MCMC) method under the model of asymmetric structured jet, in order to determine whether GRB 210731A was a good candidate produced by an asymmetric structured jet. 

\section{GRB 210731A Data}

GRB 210731A was observed by the \textit{Swift} Burst Alert Telescope \citep[BAT; ][]{Barthelmy2005SSRv} at 22:21:08 UT on July 31, 2021. We set the time when \textit{Swift} BAT being triggered as $T=0$. It has a single pulse structure at the light curve of 15–350 keV, and the duration $T_{90}$ is $22.5\pm2.8$s \citep{Stamatikos2021GCN}. $\sim$200 seconds after the BAT trigger, the \textit{Swift} X-Ray Telescope \citep[XRT,][]{Burrows2005SSRv} commenced its observation of the GRB 210731A's field and detected a prominent new X-ray source that correlated with the BAT's reported position, as documented by \citet{Troja2021GCN}. And a rapid decrease in X-ray radiation flow was observed in the following 62 seconds. The observations form 10ks to 20ks after trigger show a platform. Then, the radiation flux has continued to decrease. 

210 seconds after the trigger event, the \textit{Swift} UltraViolet and Optical Telescope \citep[UVOT,][]{Roming2005SSRv} began observing the field and capturing a single exposure of 61.7 seconds using a white filter. Starting approximately 3.27 hours after trigger, various filters were used intermittently for the following five days to continue observation \citep{Troja2021GCN,Kuin2021GCN}. 

The wide-field, robotic MeerLICHT optical telescope in Sutherland \citep{Bloemen2016SPIE} was triggered by \textit{Swift}. 286 seconds after the \textit{Swift} trigger, it began observing the BAT error circle. In its initial 60-second q-band exposure, the optical afterglow of GRB 210731A was detected. The observation lasted for approximately 4.29 hours after trigger, and continue to track the decline of afterglow in the following days. There are three peaks of comparable brightness appeared in MeerLicht's first four hours of multi-band observation \citep{deWet2023A&A}. The abnormal optical afterglow prompted a series of multi-wavelength follow-up observations, covering a spectrum ranging from X-ray to radio frequencies:

\begin{itemize}
    \item Within 4.2 hours after GRB trigger,the Gamma-Ray Burst Optical Near-Infrared Detector \citep[GROND,][]{Greiner2008PASP} which installed on the 2.2 m MPG telescope at the La Silla Observatory of the European Southern Observatory (ESO) in Chile observed the afterglow of GRB 210731A in the g', r', i',z', J, H and K bands \citep{Nicuesa2021GCN}. The subsequent observations were conducted on 1.225 days, 2.214 days, and 5.253 days \citep{Nicuesa2021GCNb}.
    \item At 9.04 hours after trigger, the 76cm Kazmann Automatic Imaging Telescope \citep[KAIT,][]{Filippenko2001ASPC} at the Lick Observatory observed afterglow in the \textit{clear} band \citep{Zheng2021GCN}.
    \item After the \textit{Swift} BAT was triggered for 1.18 days, the acquisition camera of the X-shooter spectrograph which installed on the ESO Very Large Telescope (VLT) UT3 (Meripal) obtained images in the \textit{r}, \textit{g} and \textit{z} bands. And on 1.19 days after trigger, the X-Shooter on the Very Large Telescope obtained the afterglow redshift of $z=1.2525$ \citep{Kann2021GCN}.
    \item In C and X bands, \textit{Karl G. Jansky} Very Large Array \citep[JVLA][]{Perley2011ApJ} conducted four observations from 18.2 to 118 days after trigger. In L-band, \citet{deWet2023A&A} used the MeerKAT radio telescope to observe it on 10.8, 34.1, and 59.7 days after trigger, but no afterglow was observed. They set three times the root mean square noise as the upper limit of the L-band afterglow flux.
\end{itemize}

In this work, we used the multiband data collected in \citep{deWet2023A&A}, which is shown in Figure \ref{light_curve}.

\section{Model description}
\begin{figure}
    \centering
    \includegraphics[width=8.5cm]{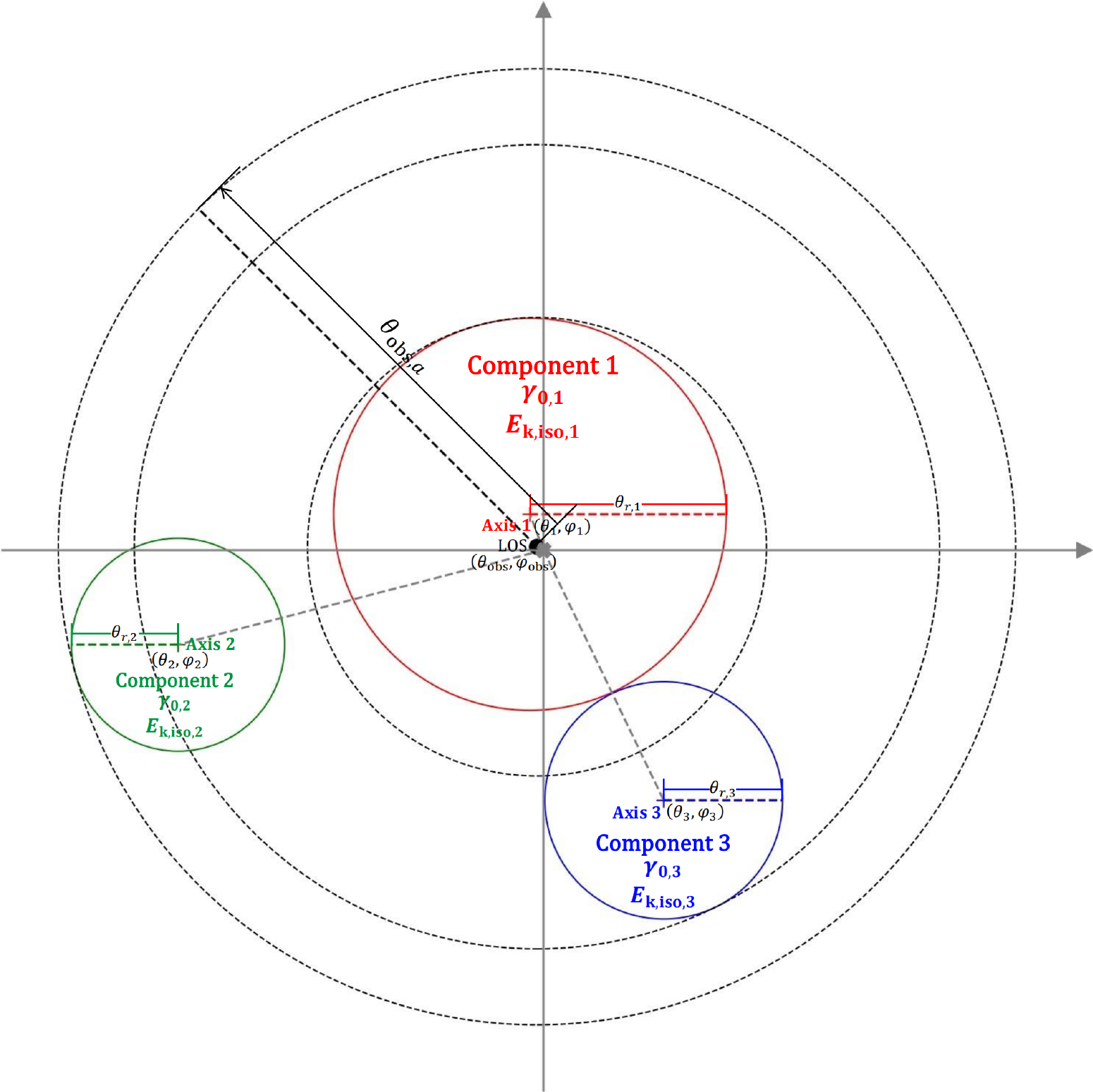}
    \caption{
    The schematic diagram represents the cross-section of an asymmetric jet. The jet consists of three components (the solid line circles with different colors), which have different initial Lorentz factor $\gamma_0$ and equivalent isotropic kinetic energy $E_{\text{k,iso}}$. We assume that different components don't have overlapping regions on the cross-section. Taking the assumed jet axis as the coordinate origin $(0,0)$, the projection of LOS on the cross-section and the axis of each component can be expressed in polar coordinates as $(\theta_{\text{obs}},\varphi_{\text{obs}})$, $(\theta_1,\varphi_1)$, $(\theta_2,\varphi_2)$, $(\theta_3,\varphi_3)$, respectively. And the half-opening angle of each component is $\theta_{r,1}$, $\theta_{r,2}$ and $\theta_{r,3}$. The point sources on the loop around the LOS on the jet (the dash line circles) have equal values of $\theta_{\text{obs},a}$.
    }
    \label{3E}
\end{figure}

\citet{Li2023MNRAS} have established theoretical descriptions of asymmetric jet structures. In order to discuss the influence of asymmetric structures, they simply assumed that the jet is uniform in the direction of the polar angle $\theta$. And a jet with an asymmetric structure can be divided into $N$ elements based on different physical parameters at different positions. Each element could be taken as a uniform "patch" and  Each "patch" can be considered independent. However, in more realistic situations, asymmetric structures are likely to be nonuniform in the direction of the polar angle $\theta$. \citet{Gill2023MNRAS} described the asymmetric structure of the jets during the GRBs' prompt emission stage with individual blobs or mini-jets (MJs) with multiple distinct properties or patches, which can be generated by a kind of hydrodynamic or hydromagnetic disturbance. It can be expected that this asymmetric structure may remain in the jet after the prompt emission stage and affect afterglow emission. Figure \ref{3E} depicts a schematic diagram of an asymmetric jet's cross-section with three components or MJs.
The Lorentz factors and equivalent isotropic kinetic energies corresponding to the three components are $\gamma_{0,1}$, $E_{\text{k,iso},1}$, $\gamma_{0,2}$, $E_{\text{k,iso},2}$ and $\gamma_{0,3}$, $E_{\text{k,iso},3}$. If we take the assumed jet axis as the origin $(0,0)$ and define the angle between a point on the jet and the axis as the polar angle $\theta$, and the circumferential direction around the jet axis as the azimuth angle $\varphi$, we can establish a polar coordinate on the cross-section of the jet (see Figure \ref{3E}). The axes of the three components are represented in polar coordinate as the corresponding polar angles and azimuth angles: $(\theta_1,\varphi_1)$, $(\theta_2,\varphi_2)$, $(\theta_3,\varphi_3)$. And the half-opening angle of each component can be represented by $\theta_{r,1}$, $\theta_{r,2}$ and $\theta_{r,3}$. Assuming there is no overlap between three components, this structure can be mathematically represented as:
\begin{equation}
    \gamma_0=\begin{cases}
    \gamma_{0,1}&\theta<\theta_{1}+\theta_{r,1}\text{cos}\left(\varphi-\varphi_1\right),\\
    \gamma_{0,2}&\theta<\theta_{2}+\theta_{r,2}\text{cos}\left(\varphi-\varphi_2\right),\\
    \gamma_{0,3}&\theta<\theta_{3}+\theta_{r,3}\text{cos}\left(\varphi-\varphi_3\right),
    \end{cases}
    \label{g0}
\end{equation}
\begin{equation}
    E_{\text{k,iso}}=\begin{cases}
    E_{\text{k,iso},1}&\theta<\theta_{1}+\theta_{r,1}\text{cos}\left(\varphi-\varphi_1\right),\\
    E_{\text{k,iso},2}&\theta<\theta_{2}+\theta_{r,2}\text{cos}\left(\varphi-\varphi_2\right),\\
    E_{\text{k,iso},3}&\theta<\theta_{3}+\theta_{r,3}\text{cos}\left(\varphi-\varphi_3\right).
    \end{cases}
    \label{E0}
\end{equation}

For the $i-$th component, the dynamic evolution is calculated based on \citet{Huang2000ApJ}. The evolution of the jet's radius $R$ over time $T$ in the frame of an on-axis observer can read as: 
\begin{equation}
    \frac{dR}{dT}=\beta c\gamma(\gamma +\sqrt{\gamma^2-1}),
    \label{eq:dR_dT}
\end{equation}
where $\gamma$ is the Lorentz factor of the jet's bulk motion, while $\beta$ is the dimensionless velocity corresponding to it. The mass of the interstellar medium swept by the component $m$ with a radius of $R$ can be described as:
\begin{equation}
    \frac{dm}{dR}=R^2\left(1-\cos{\theta_{r,i}}\right)nm_{\text{p}},
\end{equation}
where $m_{\text{p}}$ is the mass of proton. The particle number density of the interstellar medium, $n$, can be described as $AR^{-k}$, where $k$ is the wind profile variable. For the uniform interstellar medium, $k=0$. And for the stellar wind environment, $k=2$. $A$ is a constant. $\theta_{r,i}$ represents the half-opening angle of the $i-$th component. Here the lateral spread of the jet is ignored, therefore $\theta_{r,i}$ is a constant. Considering the radiation cooling effect, the evolution of jet's bulk motion Lorentz factor $\gamma$ with respect to $m$ can be written as \citep{Huang1999ChPhL,Huang1999MNRAS}:
\begin{equation}
   \frac{d\gamma}{dm}=-\frac{\gamma^2-1}{M_{\text{ej}}+\varepsilon m+2(1-\varepsilon)\gamma m},
   \label{eq:dgammma_dm}
\end{equation}
where $M_{\text{ej}}=E_0/\left(\gamma_0c^2\right)$ is the ejecta mass, and $E_0$ is the initial kinetic energy of a component. The $\varepsilon$ is radiative efficiency, which is defined as the proportion of the internal energy generated by the shock in jet's comoving frame that would be radiated, which can be described as \citep{Dai1999ApJL}:
\begin{equation}
    \varepsilon=\epsilon_e\frac{t_{\text{syn}}^{'-1}}{t_{\text{syn}}^{'-1}+t_{\text{ex}}^{'-1}},
\end{equation}
where $t^{'}_{\text{syn}}=6\pi m_ec/\left(\sigma_{\text{T}}B^{'2}\gamma_{e,\text{min}}\right)$ is the synchrotron cooling timescale, and $t^{'}_{\text{ex}}=R/\left(\gamma c\right)$ is the expansion timescale in the jet's comoving frame. The $\sigma_{\text{T}}$ stands for the cross section for Thompson scattering, and the $m_e$ stands for the mass of electron. Assuming a fraction $\epsilon_B$ as the proportion of the total internal energy generated by shock goes into the random magnetic field, thus, the magnetic energy density in the jet's comoving frame can  be estimated as
\begin{equation}
    \frac{B^{'2}}{8\pi}=\epsilon_B^2\frac{\hat{\gamma}\gamma+1}{\hat{\gamma}-1}\left(\gamma-1\right)nm_pc^2, 
\end{equation}
where $\hat{\gamma}=\left(4\gamma+1\right)/\left(3\gamma\right)$ is the adiabatic index \citep{Dai1999ApJL}. Assuming a fraction $\epsilon_e$ as the proportion of the total internal energy generated by shock goes into the electrons. And we assume that the accelerated electrons in the interstellar medium have a power-law distribution with an index of $p$ ($dN_e/d\gamma_e \propto \gamma_e^{-p}$). Thus, in the jet's comoving frame, the minimum Lorentz factor for the random motion of electrons can be derived as \citep{Huang2000ApJ}
\begin{equation}
    \gamma_{e,\text{min}}=\epsilon_e\left(\gamma-1\right)\frac{m_p\left(p-2\right)}{m_e\left(p-1\right)}+1.
\end{equation}

For synchrotron radiation, the characteristic frequency and the observed radiation power of an electron with Lorentz factor $\gamma_e$ are given by \citep{Sari1998ApJL}
\begin{equation}   \nu\left(\gamma_e\right)=\gamma\gamma_e^2\frac{q_eB}{2\pi m_ec}.
   \label{eq:nu_char}
\end{equation}
\begin{equation}   P\left(\gamma_e\right)=\frac{4}{3}\sigma_Td\gamma^2\gamma_e^2\frac{B^2}{8\pi},
   \label{eq:P_gammae}
\end{equation}
where $q_e$ is the charge of an electron. The characteristic frequency $\nu(\gamma_e)$ corresponds to the peak power of the spectrum, which can be estimated as
\begin{equation}
    P_{\nu,\text{max}}\approx\frac{P\left(\gamma_e\right)}{\nu\left(\gamma_e\right)}=\frac{m_ec^2\sigma_T}{3q_e}\gamma B.
\end{equation}
\citet{Sari1998ApJL} defined a characteristic Lorentz factor $\gamma_c$ as
\begin{equation}
    \gamma_c=\frac{6\pi m_ec}{\sigma_TB^2T}=\frac{3m_e}{16\epsilon_B\sigma_Tm_pc}\frac{1}{T\gamma^3n}.
\end{equation}
The electrons with Lorentz factors exceeding $\gamma_c$ should have significantly cooled.

The two characteristic emission frequencies $\nu_m$ and $\nu_c$ in the synchrotron spectrum are defined by the electron Lorentz factors $\gamma_{e,\text{min}}$ and $\gamma_c$. In an $n=Ar^{-k}$ environment, for the slow cooling regime ($\nu_c>\nu_m$), the self absorption frequency $\nu_a$ is
\begin{equation}
\nu_a=\begin{cases}
\left[\frac{c_1q_enR}{\left(3-k\right)B\gamma_c^5}\right]^{3/5}\nu_m &\nu_a<\nu_m,\\
\left[\frac{c_2q_enR}{\left(3-k\right)B\gamma_c^5}\right]^{2/\left(p+4\right)}\nu_m &\nu_m<\nu_a<\nu_c,\\
\left[\frac{c_2q_enR}{\left(3-k\right)B\gamma_c^5}\right]^{2/\left(p+5\right)}\left(\frac{\nu_c}{\nu_m}\right)^{1/\left(p+5\right)}\nu_m &\nu_c<\nu_a.
\end{cases}
\end{equation}
$c_1$ and $c_2$ are coefficients dependent on $p$ \citep{Wu2003MNRAS}. Assuming a point source is on the LOS, the observed flux density $F_{\nu}$ is divided into the following three situations\\
(1)$\nu_a<\nu_m<\nu_c$: 
\begin{equation}
\label{lc1}
F_{\nu}=F_{\nu,\text{max}}\begin{cases}
\left(\frac{\nu}{\nu_a}\right)^2\left(\frac{\nu_a}{\nu_m}\right)^{1/3} &\nu<\nu_a,\\
\left(\frac{\nu}{\nu_m}\right)^{1/3} &\nu_a<\nu<\nu_m,\\
\left(\frac{\nu}{\nu_m}\right)^{-\left(p-1\right)/2}F_{\nu,\text{max}} &\nu_m<\nu<\nu_c,\\
\left(\frac{\nu}{\nu_m}\right)^{-\left(p-1\right)/2}\left(\frac{\nu}{\nu_c}\right)^{-p/2} &\nu_c<\nu.
\end{cases}
\end{equation}
(2)$\nu_m<\nu_a<\nu_c$: 
\begin{equation}
\label{lc2}
F_{\nu}=F_{\nu,\text{max}}\begin{cases}
\left(\frac{\nu}{\nu_m}\right)^2\left(\frac{\nu_m}{\nu_a}\right)^{\left(p+4\right)/2}F_{\nu,\text{max}} &\nu<\nu_m,\\
\left(\frac{\nu}{\nu_a}\right)^{5/2}\left(\frac{\nu_a}{\nu_m}\right)^{-\left(p-1\right)/2}F_{\nu,\text{max}} &\nu_m<\nu<\nu_a,\\
\left(\frac{\nu}{\nu_m}\right)^{-\left(p-1\right)/2}F_{\nu,\text{max}} &\nu_a<\nu<\nu_c,\\
\left(\frac{\nu}{\nu_m}\right)^{-\left(p-1\right)/2}\left(\frac{\nu}{\nu_c}\right)^{-p/2}F_{\nu,\text{max}} &\nu_c<\nu.
\end{cases}
\end{equation}
(3)$\nu_m<\nu_c<\nu_a$: 
\begin{equation}
\label{lc3}
F_{\nu}=F_{\nu,\text{max}}\begin{cases}
\left(\frac{\nu}{\nu_m}\right)^2\left(\frac{\nu_m}{\nu_a}\right)^{\left(p+4\right)/2}\left(\frac{\nu_a}{\nu_c}\right)^{-1/2} &\nu<\nu_m,\\
\left(\frac{\nu}{\nu_a}\right)^{5/2}\left(\frac{\nu_a}{\nu_c}\right)^{-p/2}\left(\frac{\nu_c}{\nu_m}\right)^{-\left(p-1\right)/2} &\nu_m<\nu<\nu_a,\\
\left(\frac{\nu}{\nu_c}\right)^{-p/2}\left(\frac{\nu_c}{\nu_m}\right)^{-\left(p-1\right)/2} &\nu_a<\nu.
\end{cases}
\end{equation}
where $F_{\nu,\text{max}}$ stands for the peak flux density of the spectrum, which can be estimated as 
\begin{equation}
    F_{\nu,\text{max}}=\frac{N_eP_{\nu,\text{max}}}{4\pi D_L^2},
\end{equation}
where $N_e$ is the total number of swept-up electrons in the post-shock fluid. Assuming a spherical geometry, $N_e=4\pi n(R)R^3/(3-k)$. $D_L$ is the Luminosity distance from the source to the observer. And in the fast cooling regime($\nu_c<\nu_m$), the self absorption frequency $\nu_a$ is:
\begin{equation}
\nu_a=\begin{cases}
\left[\frac{c_1q_enR}{\left(3-k\right)B\gamma_c^5}\right]^{3/5}\nu_c &\nu_a<\nu_c,\\
\left[\frac{c_2q_enR}{\left(3-k\right)B\gamma_c^5}\right]^{1/3}\nu_c &\nu_c<\nu_a<\nu_m,\\
\left[\frac{c_2q_enR}{\left(3-k\right)B\gamma_c^5}\right]^{2/\left(p+5\right)}\left(\frac{\nu_m}{\nu_c}\right)^{\left(p-1\right)/\left(p+5\right)}\nu_c &\nu_m<\nu_a.
\end{cases}
\end{equation}
And the flux in fast cooling regime is\\
(1)$\nu_a<\nu_c<\nu_m$: 
\begin{equation}
\label{fc1}
F_{\nu}=F_{\nu,\text{max}}\begin{cases}
\left(\frac{\nu}{\nu_a}\right)^2\left(\frac{\nu_a}{\nu_c}\right)^{1/3} &\nu<\nu_a,\\
\left(\frac{\nu}{\nu_c}\right)^{1/3} &\nu_a<\nu<\nu_c,\\
\left(\frac{\nu}{\nu_c}\right)^{-1/2} &\nu_c<\nu<\nu_m,\\
\left(\frac{\nu_m}{\nu_c}\right)^{-1/2}\left(\frac{\nu}{\nu_m}\right)^{-p/2} &\nu_m<\nu.
\end{cases}
\end{equation}
(2)$\nu_c<\nu_a<\nu_m$: 
\begin{equation}
\label{fc2}
F_{\nu}=F_{\nu,\text{max}}\begin{cases}
\left(\frac{\nu}{\nu_c}\right)^2\left(\frac{\nu_c}{\nu_a}\right)^3 &\nu<\nu_c,\\
\left(\frac{\nu}{\nu_a}\right)^{5/2}\left(\frac{\nu_a}{\nu_c}\right)^{-1/2} &\nu_c<\nu<\nu_a,\\
\left(\frac{\nu}{\nu_c}\right)^{-1/2} &\nu_a<\nu<\nu_m,\\
\left(\frac{\nu_m}{\nu_c}\right)^{-1/2}\left(\frac{\nu}{\nu_m}\right)^{-p/2} &\nu_m<\nu.
\end{cases}
\end{equation}
(3)$\nu_c<\nu_m<\nu_a$: 
\begin{equation}
\label{fc3}
F_{\nu}=F_{\nu,\text{max}}\begin{cases}
\left(\frac{\nu}{\nu_c}\right)^2\left(\frac{\nu_c}{\nu_a}\right)^3\left(\frac{\nu_a}{\nu_m}\right)^{-\left(p-1\right)/2} &\nu<\nu_c,\\
\left(\frac{\nu}{\nu_a}\right)^{5/2}\left(\frac{\nu_a}{\nu_m}\right)^{-p/2}\left(\frac{\nu_m}{\nu_c}\right)^{-1/2} &\nu_c<\nu<\nu_a,\\
\left(\frac{\nu}{\nu_m}\right)^{-p/2}\left(\frac{\nu_m}{\nu_c}\right)^{-1/2} &\nu_a<\nu.
\end{cases}
\end{equation}

For the point sources outside the LOS, the observed flux needs to be corrected by \citep{Granot2002ApJL}
\begin{equation}
F_{\nu}=a^3F_{\nu/a}\left(at\right),
   \label{lzf}
\end{equation}
with a factor
\begin{equation}
    a=\frac{1-\beta}{1-\beta\cos{\theta_{\text{obs},a}}}\approx\frac{1}{1+\gamma^2\theta_{\text{obs},a}^2},
    \label{lz}
\end{equation}
where $\theta_{\text{obs},a}$ is the angle between LOS and the point source. All point sources with equal $\theta_{\text{obs},a}$ form a loop around the LOS. We referred to the calculation method of radiation flux proposed by  \citet{Ghisellini1999MNRAS}, and the radiation flux of the entire jet is obtained by integrating over  $\theta_{\text{obs},a}$. But only the portion of each loop within the jet contributes to the total radiation (see Figure \ref{3E}). Assuming that the proportion of the corresponding ring at $\theta_{\text{obs},a}$ within the $i-$th component is $\chi_{i}\left(\theta_{\text{obs},a}\right)/2\pi$. Overall, the total radiation flux can be calculated as 
\begin{equation}
F_{\nu}\left(t\right)=\sum\limits^n\limits_{i=1}\int_0^{\infty}a^3F_{\nu/a,i}\left(at\right)\frac{\chi_{i}\left(\theta_{\text{obs},a}\right)}{2\pi}\text{d}\theta_{\text{obs},a}. 
\label{sum}
\end{equation}

\subsection{Fitting results}
In this work, we used an asymmetric jet with 3 components to analyze multi-band observation data from multiple telescopes \citep{deWet2023A&A}. And the MCMC method was used to fit the multi-band data of GRB 210731A. We assume that the jet decelerates in uniform interstellar medium ($k=0$, the particle density $n$ is a constant).

\begin{figure*}
\centering
    \includegraphics[width=17cm]{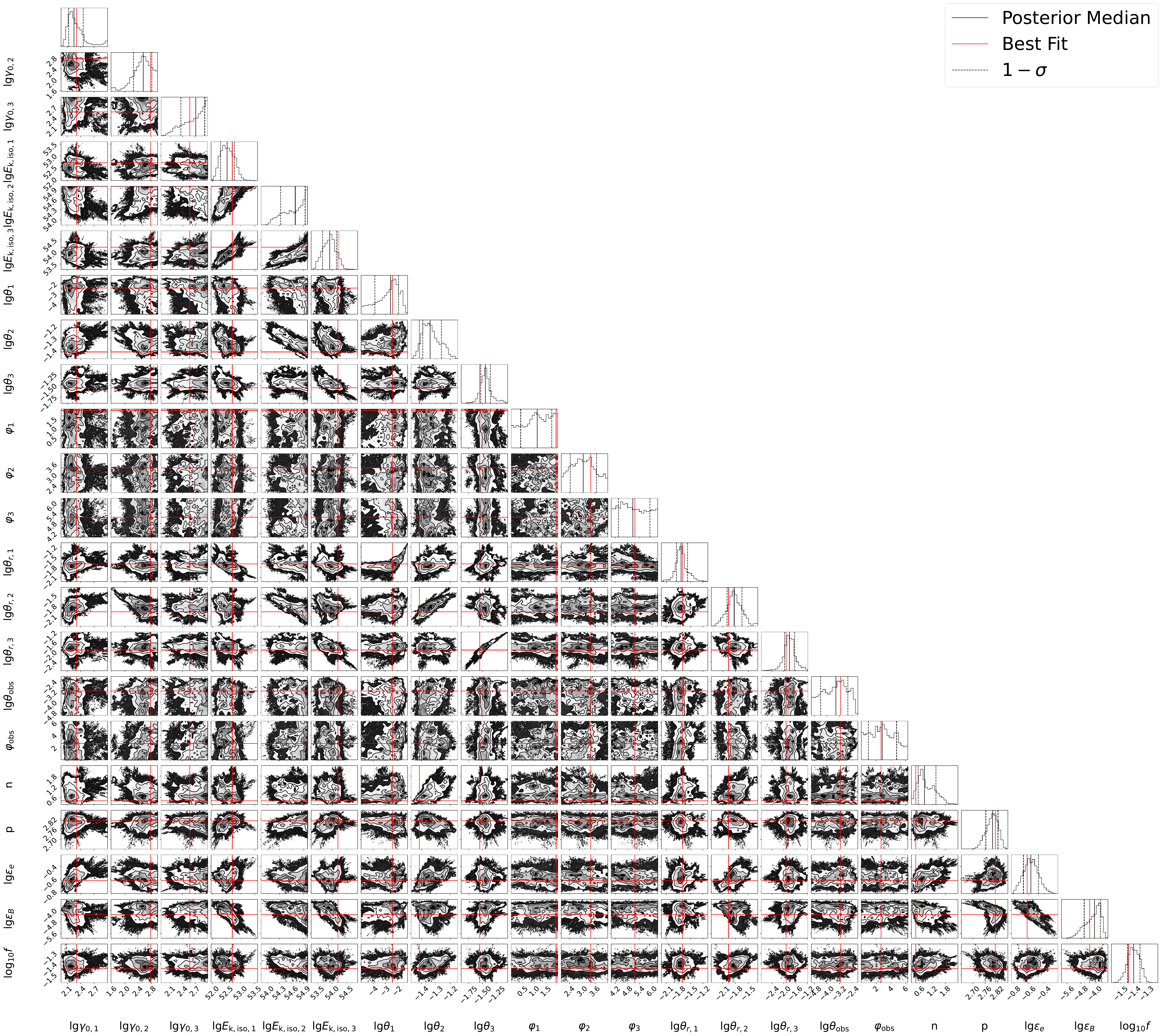}
    \caption{The fitting results of MCMC on all parameters. The best-fit values are marked with red crosses and solid red lines. The solid black lines represent the median values of the posterior distributions of the parameters. The range of $1-\sigma$ for each parameter's posterior distribution is located between two dashed lines. The fraction $f$, which underestimates the variance of the data, is used as a free parameter for fitting.}
    \label{MCMC}
\end{figure*}

\begin{figure}
    \centering
    \includegraphics[width=8.5cm]{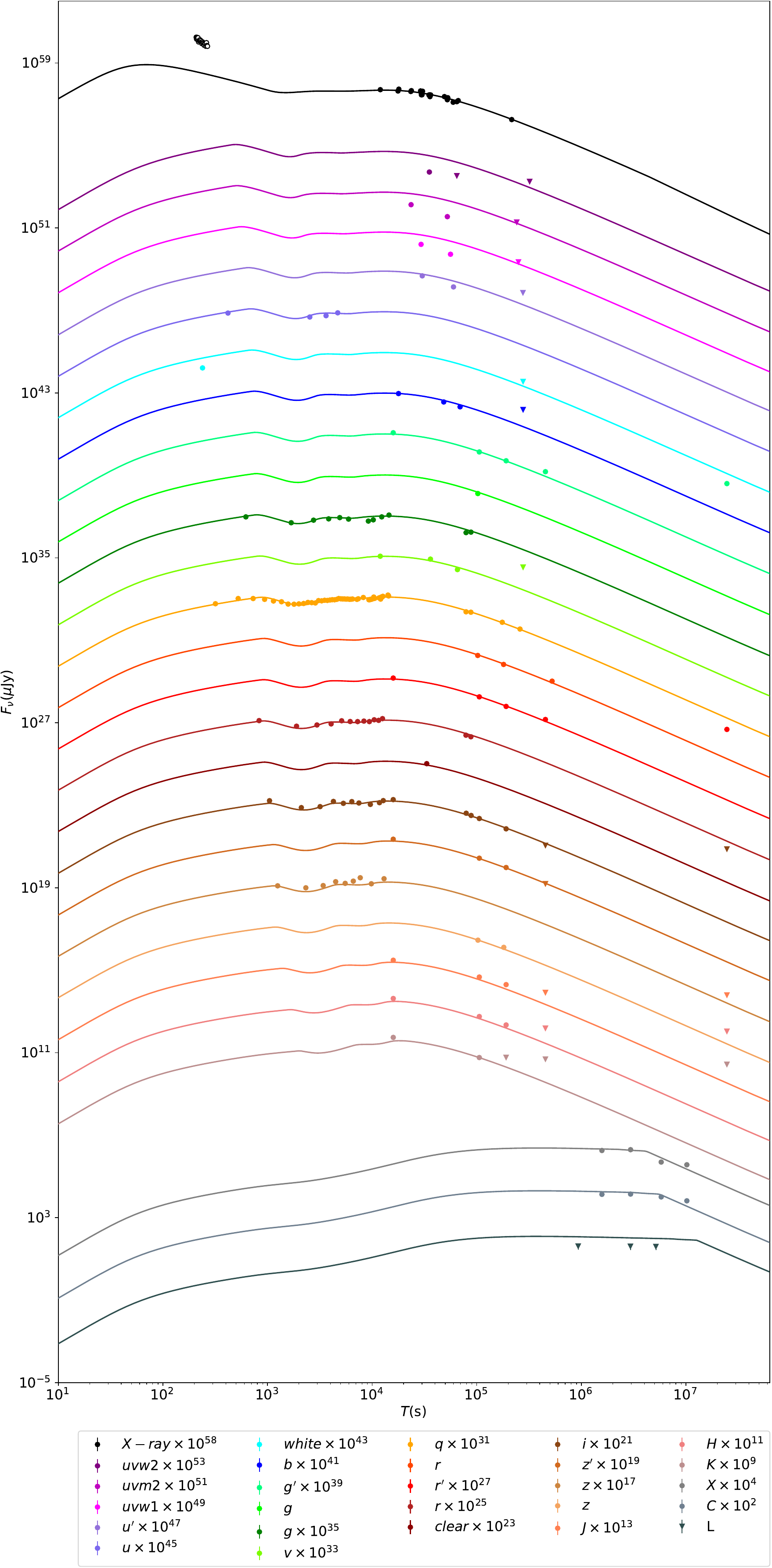}
    \caption{Multi-band afterglow data of GRB 210731A and the best fitting result. The observation data of each band is represented by points, with the inverted triangle representing the upper limit. And the hollow dots of X-ray at early time are considered as high latitude prompt emission. The theoretical light curves corresponds to the solid line.}
    \label{light_curve}
\end{figure}

\begin{figure*}
\centering
    \includegraphics[width=17cm]{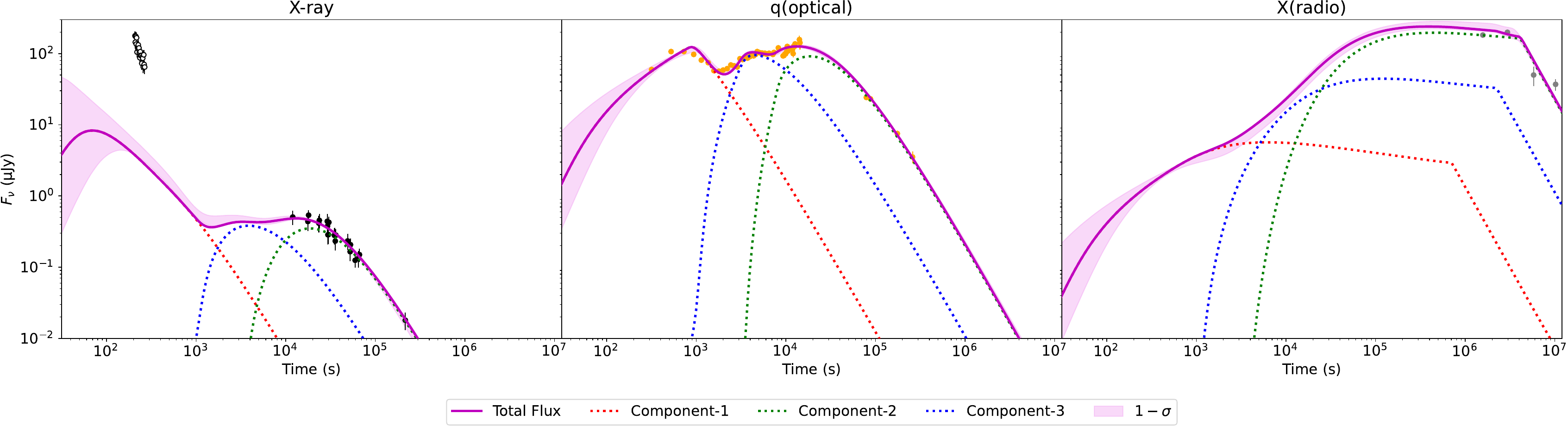}
    \caption{Same as Figure \ref{light_curve}, but we have shown the details of the X-ray, q-band, and X-band light curves. The solid line is the total flux of the jet. The $1-\sigma$ range of total radiation flux has been marked with pink shadow. And the dashed lines of different colors respectively represent the radiation of three components. The triple peaks exhibited in the optical band correspond to the radiation peaks of the three components.}
    \label{XQX}
\end{figure*}

The emcee \citep{Foreman-Mackey2013PASP} was applied to the MCMC analysis of the observation data of GRB 210731A. We used 44 walkers and 32312 steps, and found a set of parameters to fit the multi-band observation data. Our results indicate that the 3-component asymmetric jet model fits the multi-band light curve of GRB 210731A well \footnote{In the uvw1, uvm2, and uvw2 bands, the theoretical light curves are slightly higher than the observed data, which should be due to the uncertainty of extinction of ultraviolet bump in the host galaxy \citep{Corre2018A&A}. As for the data in the r-band and g-band after $10^7$s, it is consistent with the radiation of the host galaxy \citep{deWet2023A&A}.}. The best-fit values of these parameters are summarized in Table \ref{group}. As a comparison, the initial values and posterior distributions obtained from MCMC are also presented in it. Figure \ref{MCMC} shows the MCMC fitting results for all parameters. The best-fit values of these parameters and the $1-\sigma$ ranges of their posterior distributions are marked in Figure \ref{MCMC} with red solid lines and black dashed lines, respectively. The best fitting results of the light curves with respect to the multi-band afterglow data are shown as solid lines in Figure \ref{light_curve}. To better illustrate the fitting details, we separately display the fitting results for X-ray, q-band, and X-band in Figure \ref{XQX}, as well as the radiation contributions of the three components. Additionally, the $1-\sigma$ range of the light curves corresponding to the MCMC samples is depicted as pink shaded regions near the light curves of the total radiation flux. To quantitatively express the goodness of the fit, we applied the adjusted coefficient of determination, $R^2_{\text{adj}}$, to analyze the light curve corresponding to the best-fit values of these parameters \footnote{The coefficient of determination, $R^2$, quantifies how well the model explains the data by comparing the residual sum of squares, $SS_{\text{res}}$, to the total sum of squares, $SS_{\text{tot}}$, of the observed data. It is defined as $R^2=1-SS_{\text{res}}/SS_{\text{tot}}$, where $SS_{\text{tot}}$ represents the sum of the squared deviations of the observed values from their mean, reflecting the inherent variability in the data. An $R^2$ value closer to 1 indicates that more variability is explained by the model, demonstrating a better fit. To avoid inflated $R^2$ in models with many parameters, we use the adjusted coefficient of determination, $R_{\text{adj}}^2=1-(1-R^2)(N-1)/(N-P-1)$, where $N$ is the number of data points and $P$ is the number of parameters.}. Moreover, we obtained $R^2_{\text{adj}}\sim0.91$, which is close enough to 1 to indicate that our model fits the data well.

\begin{table*}
    \caption{The fitting parameters. The best-fit values are the maximum likelihood parameters from the posterior distribution of the MCMC sample. For comparison, we also present the initial settings of the MCMC and the medians of the posterior distribution with the 1$\sigma$ range of the MCMC samples.}
    \label{group}
\begin{center}
\resizebox{17cm}{!}{
\begin{tabular}{ c c c c c c}
    \hline
     & & parameters & best fit & initial & posterior \\
    \hline
    Components' parameters & Lorentz factors & $\gamma_{0,1}$ & $203.7$ & $316.2$ & $178.2_{-42.7}^{+106.9}$ \\ 
     & & $\gamma_{0,2}$ & $610.9$ & $199.5$ & $364.8_{-181.1}^{+302.0}$ \\ 
     & & $\gamma_{0,3}$ & $373.2$ & $505.8$ & $516.4_{-290.5}^{+336.7}$ \\ 
     & Isotropic kinetic energy (erg) & $E_{\text{k,iso},1}$ & $5.0\times{10^{52}}$ & $1.6\times10^{52}$ & $3.3_{-1.4}^{+2.8}\times10^{52}$ \\
     & & $E_{\text{k,iso},2}$ & $1.0\times10^{55}$ & $2.1\times10^{54}$ & $4.9_{-2.8}^{+3.7}\times10^{54}$ \\
     & & $E_{\text{k,iso},3}$ & $1.6\times10^{54}$ & $4.0\times10^{53}$ & $8.2_{-3.6}^{+5.7}\times10^{53}$ \\
     & Polar angles of axes & $\theta_1$ & $0.004$ & $0.011$ & $0.003_{-0.002}^{+0.010}$ \\
     & & $\theta_2$ & $0.038$ & $0.051$ & $0.046_{-0.005}^{+0.009}$ \\
     & & $\theta_3$ & $0.028$ & $0.035$ & $0.036_{-0.006}^{+0.008}$ \\
     & Azimuth angles of axes & $\varphi_1$ & $1.9$ & $1.0$ & $1.1_{-0.7}^{+0.6}$ \\
     & & $\varphi_2$ & $3.4$ & $3.2$ & $3.0_{-0.6}^{+0.6}$ \\
     & & $\varphi_3$ & $5.2$ & $4.9$ & $5.0_{-0.7}^{+0.8}$ \\
     & Half-opening angles & $\theta_{r,1}$ & $0.020$ & $0.023$ & $0.018_{-0.004}^{+0.008}$ \\
     & & $\theta_{r,2}$ & $0.011$ & $0.019$ & $0.015_{-0.005}^{+0.008}$ \\
     & & $\theta_{r,3}$ & $0.012$ & $0.014$ & $0.016_{-0.005}^{+0.008}$ \\
     & &  &    & & \\
    Other parameters & Shock energy into electrons & $\epsilon_e$ & $0.24$ & $0.29$ & $0.27_{-0.05}^{+0.07}$ \\
    & Shock energy into magnetic field & $\epsilon_B$ & $5.1\times10^{-5}$ & $1.6\times10^{-4}$ & $9.5_{-7.1}^{+1.1}\times10^{-5}$ \\
    & Particle number density (cm$^{-3}$) & $n$ & $0.42$ & $1.0$ & $0.82_{-0.29}^{+0.54}$ \\
    & Distribution index of electrons & $p$ & $2.8$ & $2.8$ & $2.8_{-0.03}^{+0.03}$ \\
    & View angle & $\theta_{\text{obs}}$ & $7.4\times10^{-4}$ & $3.9\times10^{-4}$ & $3.8_{-3.4}^{+18.1}\times10^{-4}$ \\
    & View azimuth & $\varphi_{\text{obs}}$ & $2.6$ & $2.4$ & $2.8_{-1.9}^{+2.0}$ \\
    & underestimated fraction of variance & $f$ & $0.036$ & $1.0$ & $0.040_{-0.004}^{+0.005}$ \\
    \hline
\end{tabular}}
\end{center}
\end{table*}

Our results indicate that the afterglow generated by the deceleration of three independent components of a jet in the uniform interstellar medium provides a good fit to the multi-band observation data. Among these three components, a higher velocity is associated with higher energy. The fastest component has a Lorentz factor exceeding $600$, with an equivalent isotropic kinetic energy approaching $10^{55}$ergs. In contrast, the slowest component has a Lorentz factor of around $200$, with an equivalent isotropic kinetic energy of $1.6\times10^{52}$ergs. The difference in Lorentz factors exceeds $400$, and their equivalent isotropic kinetic energies differ by approximately $2.3$ orders of magnitude. Given this context, the presence of a medium-speed component, characterized by a Lorentz factor of about $370$, is notable. This component is much closer to the slowest component in terms of velocity (by approximately $170$), while its energy lies between that of the other two components. Furthermore, the schematic diagram of the relative positions and sizes of the cross-sections of the three components based on the best-fit parameter values is shown in Figure \ref{3E}. The projection of LOS on the components' cross-sections is marked with a black point. We found that the LOS is localized in the component with the lowest energy and slowest speed. And the observer's LOS almost coincides with the axis of the component. As a result, the component with the lowest energy and slowest speed was observed earliest, dominating the first peak in the light curve of the optical afterglow. Among the remaining two components, the component with medium energy and speed is closer to the LOS, thus contributing to the second peak in the light curve of the optical afterglow. In contrast, the component with the highest energy and speed forms the largest angle with the LOS among the three components, causing its afterglow to be observed last and producing the final and brightest peak in the light curve of the optical afterglow. At the same time, the late-stage radiation in the X-ray and radio bands from the component with the highest energy and speed can also be well fitted to the observed data.

\section{Conclusions and Discussion}

In recent years, observations of several famous GRBs, such as the one associated with gravitational waves \citep[GRB 170817A;][]{Abbott2017PhysRevLett,Kasliwal2017Science,Xiao2017ApJL,Gao2018SCPMA,Gottlieb2018MNRAS,Lazzati2018PhRvL,Lyman2018NatAs,Piro2018ApJ,Troja2018MNRAS,Zhang2018NatureCommunications} and the one known as the “brightest of all time” \citep[GRB 221009A;][]{An2023arXiv,OConnorSci}, have suggested that GRB jets may be structured. The interpretations of the radio rebrighenting in tidal disruption events Sw J1644+57 and AT2022cmc, support that these events may contain a structured jet \citep{Wang2014,Zhou2023}. Another interesting GRB source that was recently detected, GRB 210731A, displayed multiple rebrightening features in its optical afterglow. Through MCMC fitting, we show that a multi-component asymmetric structured jet can well explain the multi-band afterglow data of GRB 210731A. 

Theoretically, there could be several potential causes for this multi-component asymmetric structured jets. For example, the presence of significant non-uniformity in the internal magnetic dissipation of a jet can lead to the development of complex and asymmetric jet structures \citep{Narayan2009MNRAS}. A more recent study conducted by \citet{Huang2019MNRAS} has demonstrated that the non-uniformity of jets can exist in the circumferential direction due to the precession of GRBs' central engine. \citet{Lamb2022Univ} used the results of the three-dimensional hydrodynamic jets in the neutron star merger environment to determine the degree of polar and rotational inhomogeneity (N × N jet model). They found that the result of these inhomogeneities in the jet's energy/Lorentz factor distribution showed some degree of rotational variation.

Distinct from the energy injection model in the stellar wind environment employed by \citet{deWet2023A&A}, we found that the multi-band observation data of the afterglow of GRB 210731A can be well fitted by the deceleration of three independent components of a jet in the uniform interstellar medium. It is noteworthy that, both in the stellar wind environment of \citet{deWet2023A&A} and in our uniform interstellar medium, the theoretical light curves exceed the upper limit of the L-band. Furthermore, corresponding to our best-fit parameters, the self-absorption frequency of the highest-energy component lies between approximately $1.7 \times 10^9$ Hz and $1.9 \times 10^9$ Hz around $10^6$ to $10^7$ seconds, which is slightly higher than the L-band. This contrasts with \citet{deWet2023A&A}'s conclusion that in a stellar wind environment, the self-absorption frequency is almost impossible to occur between the L-band and C-band. Therefore, we believe that the additional sources of opacity are required to constrain the upper limit of the L-band, similar to \citet{deWet2023A&A}'s inference, which suggests an abundance of thermal electrons in the shock region.

Compared to the energy injection model, the structured jet model does not place stringent requirements on the central engine and circumstellar environment when explaining the data of GRB 210731A. However, it is important to emphasize that the ability to distinguish between the two models is very limited based solely on the fitting of light curves, even with multi-band data fitting. To truly distinguish between these two models for similar gamma-ray bursts in the future, polarimetric observations are important. The energy injection model does not, in principle, significantly affect the polarization properties of the afterglow, while under the structured jet model, the afterglow signal will exhibit significant polarization features \citep{Lazzati2003A&A,Rossi2004MNRAS,Nakar2004ApJ,Wu2005MNRAS}, especially when the jet structure is non-axisymmetric, resulting in a significant evolution of the polarization angle \citep{Li2024submit}. Future polarimetric observations of GRB afterglows will play a crucial role in revealing the jet structure.

\section*{Acknowledgements}
This work is supported by the National Natural Science Foundation of China (Projects 12373040,12021003,12473012) and the Fundamental Research Funds for the Central Universities and the Carlsberg Foundation (CF18-0183, PI: I. Tamborra).

\bibliography{sample631}{}

\begin{thebibliography}{}
\expandafter\ifx\csname natexlab\endcsname\relax\def\natexlab#1{#1}\fi
\providecommand{\url}[1]{\href{#1}{#1}}
\providecommand{\dodoi}[1]{doi:~\href{http://doi.org/#1}{\nolinkurl{#1}}}
\providecommand{\doeprint}[1]{\href{http://ascl.net/#1}{\nolinkurl{http://ascl.net/#1}}}
\providecommand{\doarXiv}[1]{\href{https://arxiv.org/abs/#1}{\nolinkurl{https://arxiv.org/abs/#1}}}

\bibitem[{Abbott {et~al.}(2017)Abbott, Abbott, Abbott, Acernese, Ackley, Adams, Adams, Addesso, Adhikari, Adya, Affeldt, Afrough, Agarwal, Agathos, Agatsuma, Aggarwal, Aguiar, Aiello, Ain, Ajith, Allen, Allen, Allocca, Altin, Amato, Ananyeva, Anderson, Anderson, Angelova, Antier, Appert, Arai, Araya, Areeda, Arnaud, Arun, Ascenzi, Ashton, Ast, Aston, Astone, Atallah, Aufmuth, Aulbert, AultONeal, Austin, Avila-Alvarez, Babak, Bacon, Bader, Bae, Bailes, Baker, Baldaccini, Ballardin, Ballmer, Banagiri, Barayoga, Barclay, Barish, Barker, Barkett, Barone, Barr, Barsotti, Barsuglia, Barta, Barthelmy, Bartlett, Bartos, Bassiri, Basti, Batch, Bawaj, Bayley, Bazzan, B\'ecsy, Beer, Bejger, Belahcene, Bell, Berger, Bergmann, Bernuzzi, Bero, Berry, Bersanetti, Bertolini, Betzwieser, Bhagwat, Bhandare, Bilenko, Billingsley, Billman, Birch, Birney, Birnholtz, Biscans, Biscoveanu, Bisht, Bitossi, Biwer, Bizouard, Blackburn, Blackman, Blair, Blair, Blair, Bloemen, Bock, Bode, Boer, Bogaert, Bohe, Bondu, Bonilla, Bonnand,
  Boom, Bork, Boschi, Bose, Bossie, Bouffanais, Bozzi, Bradaschia, Brady, Branchesi, Brau, Briant, Brillet, Brinkmann, Brisson, Brockill, Broida, Brooks, Brown, Brown, Brunett, Buchanan, Buikema, Bulik, Bulten, Buonanno, Buskulic, Buy, Byer, Cabero, Cadonati, Cagnoli, Cahillane, Calder\'on~Bustillo, Callister, Calloni, Camp, Canepa, Canizares, Cannon, Cao, Cao, Capano, Capocasa, Carbognani, Caride, Carney, Carullo, Casanueva~Diaz, Casentini, Caudill, Cavagli\`a, Cavalier, Cavalieri, Cella, Cepeda, Cerd\'a-Dur\'an, Cerretani, Cesarini, Chamberlin, Chan, Chao, Charlton, Chase, Chassande-Mottin, Chatterjee, Chatziioannou, Cheeseboro, Chen, Chen, Chen, Cheng, Chia, Chincarini, Chiummo, Chmiel, Cho, Cho, Chow, Christensen, Chu, Chua, Chua, Chung, Chung, Ciani, Ciolfi, Cirelli, Cirone, Clara, Clark, Clearwater, Cleva, Cocchieri, Coccia, Cohadon, Cohen, Colla, Collette, Cominsky, Constancio, Conti, Cooper, Corban, Corbitt, Cordero-Carri\'on, Corley, Cornish, Corsi, Cortese, Costa, Coughlin, Coughlin, Coulon,
  Countryman, Couvares, Covas, Cowan, Coward, Cowart, Coyne, Coyne, Creighton, Creighton, Cripe, Crowder, Cullen, Cumming, Cunningham, Cuoco, Dal~Canton, D\'alya, Danilishin, D'Antonio, Danzmann, Dasgupta, Da~Silva~Costa, Dattilo, Dave, Davier, Davis, Daw, Day, De, DeBra, Degallaix, De~Laurentis, Del\'eglise, Del~Pozzo, Demos, Denker, Dent, De~Pietri, Dergachev, De~Rosa, DeRosa, De~Rossi, DeSalvo, de~Varona, Devenson, Dhurandhar, D\'{\i}az, Dietrich, Di~Fiore, Di~Giovanni, Di~Girolamo, Di~Lieto, Di~Pace, Di~Palma, Di~Renzo, Doctor, Dolique, Donovan, Dooley, Doravari, Dorrington, Douglas, Dovale~\'Alvarez, Downes, Drago, Dreissigacker, Driggers, Du, Ducrot, Dudi, Dupej, Dwyer, Edo, Edwards, Effler, Eggenstein, Ehrens, Eichholz, Eikenberry, Eisenstein, Essick, Estevez, Etienne, Etzel, Evans, Evans, Factourovich, Fafone, Fair, Fairhurst, Fan, Farinon, Farr, Farr, Fauchon-Jones, Favata, Fays, Fee, Fehrmann, Feicht, Fejer, Fernandez-Galiana, Ferrante, Ferreira, Ferrini, Fidecaro, Finstad, Fiori, Fiorucci,
  Fishbach, Fisher, Fitz-Axen, Flaminio, Fletcher, Fong, Font, Forsyth, Forsyth, Fournier, Frasca, Frasconi, Frei, Freise, Frey, Frey, Fries, Fritschel, Frolov, Fulda, Fyffe, Gabbard, Gadre, Gaebel, Gair, Gammaitoni, Ganija, Gaonkar, Garcia-Quiros, Garufi, Gateley, Gaudio, Gaur, Gayathri, Gehrels, Gemme, Genin, Gennai, George, George, Gergely, Germain, Ghonge, Ghosh, Ghosh, Ghosh, Giaime, Giardina, Giazotto, Gill, Glover, Goetz, Goetz, Gomes, Goncharov, Gonz\'alez, Gonzalez~Castro, Gopakumar, Gorodetsky, Gossan, Gosselin, Gouaty, Grado, Graef, Granata, Grant, Gras, Gray, Greco, Green, Gretarsson, Groot, Grote, Grunewald, Gruning, Guidi, Guo, Gupta, Gupta, Gushwa, Gustafson, Gustafson, Halim, Hall, Hall, Hamilton, Hammond, Haney, Hanke, Hanks, Hanna, Hannam, Hannuksela, Hanson, Hardwick, Harms, Harry, Harry, Hart, Haster, Haughian, Healy, Heidmann, Heintze, Heitmann, Hello, Hemming, Hendry, Heng, Hennig, Heptonstall, Heurs, Hild, Hinderer, Ho, Hoak, Hofman, Holt, Holz, Hopkins, Horst, Hough, Houston, Howell,
  Hreibi, Hu, Huerta, Huet, Hughey, Husa, Huttner, Huynh-Dinh, Indik, Inta, Intini, Isa, Isac, Isi, Iyer, Izumi, Jacqmin, Jani, Jaranowski, Jawahar, Jim\'enez-Forteza, Johnson, Johnson-McDaniel, Jones, Jones, Jonker, Ju, Junker, Kalaghatgi, Kalogera, Kamai, Kandhasamy, Kang, Kanner, Kapadia, Karki, Karvinen, Kasprzack, Kastaun, Katolik, Katsavounidis, Katzman, Kaufer, Kawabe, K\'ef\'elian, Keitel, Kemball, Kennedy, Kent, Key, Khalili, Khan, Khan, Khan, Khazanov, Kijbunchoo, Kim, Kim, Kim, Kim, Kim, Kim, Kimbrell, King, King, Kinley-Hanlon, Kirchhoff, Kissel, Kleybolte, Klimenko, Knowles, Koch, Koehlenbeck, Koley, Kondrashov, Kontos, Korobko, Korth, Kowalska, Kozak, Kr\"amer, Kringel, Krishnan, Kr\'olak, Kuehn, Kumar, Kumar, Kumar, Kuo, Kutynia, Kwang, Lackey, Lai, Landry, Lang, Lange, Lantz, Lanza, Larson, Lartaux-Vollard, Lasky, Laxen, Lazzarini, Lazzaro, Leaci, Leavey, Lee, Lee, Lee, Lee, Lee, Lehmann, Lenon, Leon, Leonardi, Leroy, Letendre, Levin, Li, Linker, Littenberg, Liu, Liu, Lo, Lockerbie, London,
  Lord, Lorenzini, Loriette, Lormand, Losurdo, Lough, Lousto, Lovelace, L\"uck, Lumaca, Lundgren, Lynch, Ma, Macas, Macfoy, Machenschalk, MacInnis, Macleod, Maga\~na Hernandez, Maga\~na Sandoval, Maga\~na Zertuche, Magee, Majorana, Maksimovic, Man, Mandic, Mangano, Mansell, Manske, Mantovani, Marchesoni, Marion, M\'arka, M\'arka, Markakis, Markosyan, Markowitz, Maros, Marquina, Marsh, Martelli, Martellini, Martin, Martin, Martynov, Marx, Mason, Massera, Masserot, Massinger, Masso-Reid, Mastrogiovanni, Matas, Matichard, Matone, Mavalvala, Mazumder, McCarthy, McClelland, McCormick, McCuller, McGuire, McIntyre, McIver, McManus, McNeill, McRae, McWilliams, Meacher, Meadors, Mehmet, Meidam, Mejuto-Villa, Melatos, Mendell, Mercer, Merilh, Merzougui, Meshkov, Messenger, Messick, Metzdorff, Meyers, Miao, Michel, Middleton, Mikhailov, Milano, Miller, Miller, Miller, Millhouse, Milovich-Goff, Minazzoli, Minenkov, Ming, Mishra, Mitra, Mitrofanov, Mitselmakher, Mittleman, Moffa, Moggi, Mogushi, Mohan, Mohapatra, Molina,
  Montani, Moore, Moraru, Moreno, Morisaki, Morriss, Mours, Mow-Lowry, Mueller, Muir, Mukherjee, Mukherjee, Mukherjee, Mukund, Mullavey, Munch, Mu\~niz, Muratore, Murray, Nagar, Napier, Nardecchia, Naticchioni, Nayak, Neilson, Nelemans, Nelson, Nery, Neunzert, Nevin, Newport, Newton, Ng, Nguyen, Nguyen, Nichols, Nielsen, Nissanke, Nitz, Noack, Nocera, Nolting, North, Nuttall, Oberling, O'Dea, Ogin, Oh, Oh, Ohme, Okada, Oliver, Oppermann, Oram, O'Reilly, Ormiston, Ortega, O'Shaughnessy, Ossokine, Ottaway, Overmier, Owen, Pace, Page, Page, Pai, Pai, Palamos, Palashov, Palomba, Pal-Singh, Pan, Pan, Pang, Pang, Pankow, Pannarale, Pant, Paoletti, Paoli, Papa, Parida, Parker, Pascucci, Pasqualetti, Passaquieti, Passuello, Patil, Patricelli, Pearlstone, Pedraza, Pedurand, Pekowsky, Pele, Penn, Perez, Perreca, Perri, Pfeiffer, Phelps, Piccinni, Pichot, Piergiovanni, Pierro, Pillant, Pinard, Pinto, Pirello, Pitkin, Poe, Poggiani, Popolizio, Porter, Post, Powell, Prasad, Pratt, Pratten, Predoi, Prestegard, Prijatelj,
  Principe, Privitera, Prix, Prodi, Prokhorov, Puncken, Punturo, Puppo, P\"urrer, Qi, Quetschke, Quintero, Quitzow-James, Raab, Rabeling, Radkins, Raffai, Raja, Rajan, Rajbhandari, Rakhmanov, Ramirez, Ramos-Buades, Rapagnani, Raymond, Razzano, Read, Regimbau, Rei, Reid, Reitze, Ren, Reyes, Ricci, Ricker, Rieger, Riles, Rizzo, Robertson, Robie, Robinet, Rocchi, Rolland, Rollins, Roma, Romano, Romano, Romel, Romie, Rosi\ifmmode~\acute{n}\else \'{n}\fi{}ska, Ross, Rowan, R\"udiger, Ruggi, Rutins, Ryan, Sachdev, Sadecki, Sadeghian, Sakellariadou, Salconi, Saleem, Salemi, Samajdar, Sammut, Sampson, Sanchez, Sanchez, Sanchis-Gual, Sandberg, Sanders, Sassolas, Sathyaprakash, Saulson, Sauter, Savage, Sawadsky, Schale, Scheel, Scheuer, Schmidt, Schmidt, Schnabel, Schofield, Sch\"onbeck, Schreiber, Schuette, Schulte, Schutz, Schwalbe, Scott, Scott, Seidel, Sellers, Sengupta, Sentenac, Sequino, Sergeev, Shaddock, Shaffer, Shah, Shahriar, Shaner, Shao, Shapiro, Shawhan, Sheperd, Shoemaker, Shoemaker, Siellez, Siemens,
  Sieniawska, Sigg, Silva, Singer, Singh, Singhal, Sintes, Slagmolen, Smith, Smith, Smith, Somala, Son, Sonnenberg, Sorazu, Sorrentino, Souradeep, Spencer, Srivastava, Staats, Staley, Steinke, Steinlechner, Steinlechner, Steinmeyer, Stevenson, Stone, Stops, Strain, Stratta, Strigin, Strunk, Sturani, Stuver, Summerscales, Sun, Sunil, Suresh, Sutton, Swinkels, Szczepa\ifmmode~\acute{n}\else \'{n}\fi{}czyk, Tacca, Tait, Talbot, Talukder, Tanner, T\'apai, Taracchini, Tasson, Taylor, Taylor, Tewari, Theeg, Thies, Thomas, Thomas, Thomas, Thorne, Thorne, Thrane, Tiwari, Tiwari, Tokmakov, Toland, Tonelli, Tornasi, Torres-Forn\'e, Torrie, T\"oyr\"a, Travasso, Traylor, Trinastic, Tringali, Trozzo, Tsang, Tse, Tso, Tsukada, Tsuna, Tuyenbayev, Ueno, Ugolini, Unnikrishnan, Urban, Usman, Vahlbruch, Vajente, Valdes, Vallisneri, van Bakel, van Beuzekom, van~den Brand, Van Den~Broeck, Vander-Hyde, van~der Schaaf, van Heijningen, van Veggel, Vardaro, Varma, Vass, Vas\'uth, Vecchio, Vedovato, Veitch, Veitch, Venkateswara,
  Venugopalan, Verkindt, Vetrano, Vicer\'e, Viets, Vinciguerra, Vine, Vinet, Vitale, Vo, Vocca, Vorvick, Vyatchanin, Wade, Wade, Wade, Walet, Walker, Wallace, Walsh, Wang, Wang, Wang, Wang, Wang, Ward, Warner, Was, Watchi, Weaver, Wei, Weinert, Weinstein, Weiss, Wen, Wessel, We\ss{}els, Westerweck, Westphal, Wette, Whelan, Whitcomb, Whiting, Whittle, Wilken, Williams, Williams, Williamson, Willis, Willke, Wimmer, Winkler, Wipf, Wittel, Woan, Woehler, Wofford, Wong, Worden, Wright, Wu, Wysocki, Xiao, Yamamoto, Yancey, Yang, Yap, Yazback, Yu, Yu, Yvert, Zadro\ifmmode~\dot{z}\else \.{z}\fi{}ny, Zanolin, Zelenova, Zendri, Zevin, Zhang, Zhang, Zhang, Zhang, Zhao, Zhou, Zhou, Zhu, Zhu, Zimmerman, Zucker, \& Zweizig}]{Abbott2017PhysRevLett}
Abbott, B.~P., Abbott, R., Abbott, T.~D., {et~al.} 2017, Phys. Rev. Lett., 119, 161101, \dodoi{10.1103/PhysRevLett.119.161101}

\bibitem[{{An} {et~al.}(2023){An}, {Antier}, {Bi}, {Bu}, {Cai}, {Cao}, {Camisasca}, {Chang}, {Chen}, {Chen}, {Chen}, {Chen}, {Chen}, {Chen}, {Chen}, {Coughlin}, {Cui}, {Dai}, {Hussenot-Desenonges}, {Du}, {Du}, {Du}, {Fan}, {Frontera}, {Gao}, {Gao}, {Ge}, {Gong}, {Gu}, {Guan}, {Guo}, {Guo}, {Guidorzi}, {Han}, {He}, {He}, {Hou}, {Huang}, {Huo}, {Ji}, {Jia}, {Jiang}, {Kann}, {Klotz}, {Kong}, {Lan}, {Li}, {Li}, {Li}, {Li}, {Li}, {Li}, {Li}, {Li}, {Li}, {Li}, {Li}, {Li}, {Li}, {Liang}, {Liang}, {Liao}, {Lin}, {Liu}, {Liu}, {Liu}, {Liu}, {Liu}, {Liu}, {Liu}, {Lu}, {Lu}, {Lu}, {Luo}, {Luo}, {Ma}, {Ma}, {Ma}, {Ma}, {Maccary}, {Mao}, {Meng}, {Nie}, {Orlandini}, {Ou}, {Peng}, {Peng}, {Qiao}, {Qu}, {Ren}, {Shi}, {Shi}, {Song}, {Song}, {Su}, {Sun}, {Sun}, {Sun}, {Tan}, {Tan}, {Tao}, {Tuo}, {Turpin}, {Wang}, {Wang}, {Wang}, {Wang}, {Wang}, {Wang}, {Wang}, {Wang}, {Wang}, {Wang}, {Wang}, {Wang}, {Wang}, {Wang}, {Wen}, {Wu}, {Wu}, {Wu}, {Xiao}, {Xiao}, {Xiao}, {Xie}, {Xiong}, {Xiong}, {Xu}, {Xu}, {Xu}, {Xu}, {Xu}, {Xu},
  {Xue}, {Yang}, {Yang}, {Yang}, {Ye}, {Yi}, {Yi}, {Yin}, {You}, {Yu}, {Yu}, {Yu}, {Zeng}, {Zhang}, {Zhang}, {Zhang}, {Zhang}, {Zhang}, {Zhang}, {Zhang}, {Zhang}, {Zhang}, {Zhang}, {Zhang}, {Zhang}, {Zhang}, {Zhang}, {Zhang}, {Zhang}, {Zhang}, {Zhang}, {Zhang}, {Zhao}, {Zhao}, {Zhao}, {Zhao}, {Zhao}, {Zhao}, {Zhao}, {Zhao}, {Zheng}, {Zheng}, {Zhou}, {Zhou}, \& {Zhu}}]{An2023arXiv}
{An}, Z.-H., {Antier}, S., {Bi}, X.-Z., {et~al.} 2023, arXiv e-prints, arXiv:2303.01203, \dodoi{10.48550/arXiv.2303.01203}

\bibitem[{{Barthelmy} {et~al.}(2005){Barthelmy}, {Barbier}, {Cummings}, {Fenimore}, {Gehrels}, {Hullinger}, {Krimm}, {Markwardt}, {Palmer}, {Parsons}, {Sato}, {Suzuki}, {Takahashi}, {Tashiro}, \& {Tueller}}]{Barthelmy2005SSRv}
{Barthelmy}, S.~D., {Barbier}, L.~M., {Cummings}, J.~R., {et~al.} 2005, \ssr, 120, 143, \dodoi{10.1007/s11214-005-5096-3}

\bibitem[{{Beniamini} {et~al.}(2020){Beniamini}, {Granot}, \& {Gill}}]{Beniamini2020MNRAS}
{Beniamini}, P., {Granot}, J., \& {Gill}, R. 2020, \mnras, 493, 3521, \dodoi{10.1093/mnras/staa538}

\bibitem[{{Bloemen} {et~al.}(2016){Bloemen}, {Groot}, {Woudt}, {Klein Wolt}, {McBride}, {Nelemans}, {K{\"o}rding}, {Pretorius}, {Roelfsema}, {Bettonvil}, {Balster}, {Bakker}, {Dolron}, {van Elteren}, {Elswijk}, {Engels}, {Fender}, {Fokker}, {de Haan}, {Hagoort}, {de Hoog}, {ter Horst}, {van der Kevie}, {Koz{\l}owski}, {Kragt}, {Lech}, {Le Poole}, {Lesman}, {Morren}, {Navarro}, {Paalberends}, {Paterson}, {Paw{\l}aszek}, {Pessemier}, {Raskin}, {Rutten}, {Scheers}, {Schuil}, \& {Sybilski}}]{Bloemen2016SPIE}
{Bloemen}, S., {Groot}, P., {Woudt}, P., {et~al.} 2016, in Society of Photo-Optical Instrumentation Engineers (SPIE) Conference Series, Vol. 9906, Ground-based and Airborne Telescopes VI, ed. H.~J. {Hall}, R.~{Gilmozzi}, \& H.~K. {Marshall}, 990664, \dodoi{10.1117/12.2232522}

\bibitem[{{Burrows} {et~al.}(2005){Burrows}, {Hill}, {Nousek}, {Kennea}, {Wells}, {Osborne}, {Abbey}, {Beardmore}, {Mukerjee}, {Short}, {Chincarini}, {Campana}, {Citterio}, {Moretti}, {Pagani}, {Tagliaferri}, {Giommi}, {Capalbi}, {Tamburelli}, {Angelini}, {Cusumano}, {Br{\"a}uninger}, {Burkert}, \& {Hartner}}]{Burrows2005SSRv}
{Burrows}, D.~N., {Hill}, J.~E., {Nousek}, J.~A., {et~al.} 2005, \ssr, 120, 165, \dodoi{10.1007/s11214-005-5097-2}

\bibitem[{{Corre} {et~al.}(2018){Corre}, {Buat}, {Basa}, {Boissier}, {Japelj}, {Palmerio}, {Salvaterra}, {Vergani}, \& {Zafar}}]{Corre2018A&A}
{Corre}, D., {Buat}, V., {Basa}, S., {et~al.} 2018, \aap, 617, A141, \dodoi{10.1051/0004-6361/201832926}

\bibitem[{{Costa} {et~al.}(1997){Costa}, {Frontera}, {Heise}, {Feroci}, {in't Zand}, {Fiore}, {Cinti}, {Dal Fiume}, {Nicastro}, {Orlandini}, {Palazzi}, {Rapisarda\#}, {Zavattini}, {Jager}, {Parmar}, {Owens}, {Molendi}, {Cusumano}, {Maccarone}, {Giarrusso}, {Coletta}, {Antonelli}, {Giommi}, {Muller}, {Piro}, \& {Butler}}]{Costa1997Natur}
{Costa}, E., {Frontera}, F., {Heise}, J., {et~al.} 1997, \nat, 387, 783, \dodoi{10.1038/42885}

\bibitem[{{Dai} \& {Lu}(1998)}]{Dai1998A&A}
{Dai}, Z.~G., \& {Lu}, T. 1998, \aap, 333, L87, \dodoi{10.48550/arXiv.astro-ph/9810402}

\bibitem[{{Dai} \& {Lu}(1999)}]{Dai1999ApJL}
---. 1999, \apjl, 519, L155, \dodoi{10.1086/312127}

\bibitem[{{de Wet} {et~al.}(2023){de Wet}, {Laskar}, {Groot}, {Cavallaro}, {Nicuesa Guelbenzu}, {Chastain}, {Izzo}, {Levan}, {Malesani}, {Monageng}, {van der Horst}, {Zheng}, {Bloemen}, {Filippenko}, {Kann}, {Klose}, {Pieterse}, {Rau}, {Vreeswijk}, {Woudt}, \& {Zhu}}]{deWet2023A&A}
{de Wet}, S., {Laskar}, T., {Groot}, P.~J., {et~al.} 2023, \aap, 671, A116, \dodoi{10.1051/0004-6361/202244917}

\bibitem[{{Filippenko} {et~al.}(2001){Filippenko}, {Li}, {Treffers}, \& {Modjaz}}]{Filippenko2001ASPC}
{Filippenko}, A.~V., {Li}, W.~D., {Treffers}, R.~R., \& {Modjaz}, M. 2001, in Astronomical Society of the Pacific Conference Series, Vol. 246, IAU Colloq. 183: Small Telescope Astronomy on Global Scales, ed. B.~{Paczynski}, W.-P. {Chen}, \& C.~{Lemme}, 121

\bibitem[{{Foreman-Mackey} {et~al.}(2013){Foreman-Mackey}, {Hogg}, {Lang}, \& {Goodman}}]{Foreman-Mackey2013PASP}
{Foreman-Mackey}, D., {Hogg}, D.~W., {Lang}, D., \& {Goodman}, J. 2013, \pasp, 125, 306, \dodoi{10.1086/670067}

\bibitem[{{Frail} {et~al.}(1997){Frail}, {Kulkarni}, {Nicastro}, {Feroci}, \& {Taylor}}]{Frail1997Natur}
{Frail}, D.~A., {Kulkarni}, S.~R., {Nicastro}, L., {Feroci}, M., \& {Taylor}, G.~B. 1997, \nat, 389, 261, \dodoi{10.1038/38451}

\bibitem[{{Gao}(2018)}]{Gao2018SCPMA}
{Gao}, H. 2018, Science China Physics, Mechanics, and Astronomy, 61, 59531, \dodoi{10.1007/s11433-017-9149-3}

\bibitem[{{Gao} {et~al.}(2013){Gao}, {Lei}, {Zou}, {Wu}, \& {Zhang}}]{Gao2013NewAR}
{Gao}, H., {Lei}, W.-H., {Zou}, Y.-C., {Wu}, X.-F., \& {Zhang}, B. 2013, \nar, 57, 141, \dodoi{10.1016/j.newar.2013.10.001}

\bibitem[{{Gehrels} {et~al.}(2004){Gehrels}, {Chincarini}, {Giommi}, {Mason}, {Nousek}, {Wells}, {White}, {Barthelmy}, {Burrows}, {Cominsky}, {Hurley}, {Marshall}, {M{\'e}sz{\'a}ros}, {Roming}, {Angelini}, {Barbier}, {Belloni}, {Campana}, {Caraveo}, {Chester}, {Citterio}, {Cline}, {Cropper}, {Cummings}, {Dean}, {Feigelson}, {Fenimore}, {Frail}, {Fruchter}, {Garmire}, {Gendreau}, {Ghisellini}, {Greiner}, {Hill}, {Hunsberger}, {Krimm}, {Kulkarni}, {Kumar}, {Lebrun}, {Lloyd-Ronning}, {Markwardt}, {Mattson}, {Mushotzky}, {Norris}, {Osborne}, {Paczynski}, {Palmer}, {Park}, {Parsons}, {Paul}, {Rees}, {Reynolds}, {Rhoads}, {Sasseen}, {Schaefer}, {Short}, {Smale}, {Smith}, {Stella}, {Tagliaferri}, {Takahashi}, {Tashiro}, {Townsley}, {Tueller}, {Turner}, {Vietri}, {Voges}, {Ward}, {Willingale}, {Zerbi}, \& {Zhang}}]{Gehrels2004ApJ}
{Gehrels}, N., {Chincarini}, G., {Giommi}, P., {et~al.} 2004, \apj, 611, 1005, \dodoi{10.1086/422091}

\bibitem[{{Ghisellini} \& {Lazzati}(1999)}]{Ghisellini1999MNRAS}
{Ghisellini}, G., \& {Lazzati}, D. 1999, \mnras, 309, L7, \dodoi{10.1046/j.1365-8711.1999.03025.x}

\bibitem[{{Gill} \& {Granot}(2023)}]{Gill2023MNRAS}
{Gill}, R., \& {Granot}, J. 2023, \mnras, \dodoi{10.1093/mnras/stad3991}

\bibitem[{{Gottlieb} {et~al.}(2021){Gottlieb}, {Nakar}, \& {Bromberg}}]{Gottlieb2021MNRAS}
{Gottlieb}, O., {Nakar}, E., \& {Bromberg}, O. 2021, \mnras, 500, 3511, \dodoi{10.1093/mnras/staa3501}

\bibitem[{{Gottlieb} {et~al.}(2018){Gottlieb}, {Nakar}, {Piran}, \& {Hotokezaka}}]{Gottlieb2018MNRAS}
{Gottlieb}, O., {Nakar}, E., {Piran}, T., \& {Hotokezaka}, K. 2018, \mnras, 479, 588, \dodoi{10.1093/mnras/sty1462}

\bibitem[{{Granot} {et~al.}(2002){Granot}, {Panaitescu}, {Kumar}, \& {Woosley}}]{Granot2002ApJL}
{Granot}, J., {Panaitescu}, A., {Kumar}, P., \& {Woosley}, S.~E. 2002, \apjl, 570, L61, \dodoi{10.1086/340991}

\bibitem[{{Greiner} {et~al.}(2008){Greiner}, {Bornemann}, {Clemens}, {Deuter}, {Hasinger}, {Honsberg}, {Huber}, {Huber}, {Krauss}, {Kr{\"u}hler}, {K{\"u}pc{\"u} Yolda{\c{s}}}, {Mayer-Hasselwander}, {Mican}, {Primak}, {Schrey}, {Steiner}, {Szokoly}, {Th{\"o}ne}, {Yolda{\c{s}}}, {Klose}, {Laux}, \& {Winkler}}]{Greiner2008PASP}
{Greiner}, J., {Bornemann}, W., {Clemens}, C., {et~al.} 2008, \pasp, 120, 405, \dodoi{10.1086/587032}

\bibitem[{{Greiner} {et~al.}(2009){Greiner}, {Kr{\"u}hler}, {McBreen}, {Ajello}, {Giannios}, {Schwarz}, {Savaglio}, {Yolda{\c{s}}}, {Clemens}, {Stefanescu}, {Sala}, {Bertoldi}, {Szokoly}, \& {Klose}}]{Greiner2009ApJ}
{Greiner}, J., {Kr{\"u}hler}, T., {McBreen}, S., {et~al.} 2009, \apj, 693, 1912, \dodoi{10.1088/0004-637X/693/2/1912}

\bibitem[{{Huang} {et~al.}(2019){Huang}, {Lin}, {Liu}, {Ren}, {Wang}, {Liu}, \& {Liang}}]{Huang2019MNRAS}
{Huang}, B.-Q., {Lin}, D.-B., {Liu}, T., {et~al.} 2019, \mnras, 487, 3214, \dodoi{10.1093/mnras/stz1426}

\bibitem[{{Huang} {et~al.}(1999{\natexlab{a}}){Huang}, {Dai}, \& {Lu}}]{Huang1999ChPhL}
{Huang}, Y.-f., {Dai}, Z.-g., \& {Lu}, T. 1999{\natexlab{a}}, Chinese Physics Letters, 16, 775, \dodoi{10.1088/0256-307X/16/10/027}

\bibitem[{{Huang} {et~al.}(1999{\natexlab{b}}){Huang}, {Dai}, \& {Lu}}]{Huang1999MNRAS}
{Huang}, Y.~F., {Dai}, Z.~G., \& {Lu}, T. 1999{\natexlab{b}}, \mnras, 309, 513, \dodoi{10.1046/j.1365-8711.1999.02887.x}

\bibitem[{{Huang} {et~al.}(2000){Huang}, {Gou}, {Dai}, \& {Lu}}]{Huang2000ApJ}
{Huang}, Y.~F., {Gou}, L.~J., {Dai}, Z.~G., \& {Lu}, T. 2000, \apj, 543, 90, \dodoi{10.1086/317076}

\bibitem[{{Huang} {et~al.}(2004){Huang}, {Wu}, {Dai}, {Ma}, \& {Lu}}]{Huang2004ApJ}
{Huang}, Y.~F., {Wu}, X.~F., {Dai}, Z.~G., {Ma}, H.~T., \& {Lu}, T. 2004, \apj, 605, 300, \dodoi{10.1086/382202}

\bibitem[{{Kann} {et~al.}(2021){Kann}, {Izzo}, {Levan}, {Malesani}, {de Wet}, \& {Stargate Collaboration}}]{Kann2021GCN}
{Kann}, D.~A., {Izzo}, L., {Levan}, A.~J., {et~al.} 2021, GRB Coordinates Network, 30583, 1

\bibitem[{{Kasliwal} {et~al.}(2017){Kasliwal}, {Nakar}, {Singer}, {Kaplan}, {Cook}, {Van Sistine}, {Lau}, {Fremling}, {Gottlieb}, {Jencson}, {Adams}, {Feindt}, {Hotokezaka}, {Ghosh}, {Perley}, {Yu}, {Piran}, {Allison}, {Anupama}, {Balasubramanian}, {Bannister}, {Bally}, {Barnes}, {Barway}, {Bellm}, {Bhalerao}, {Bhattacharya}, {Blagorodnova}, {Bloom}, {Brady}, {Cannella}, {Chatterjee}, {Cenko}, {Cobb}, {Copperwheat}, {Corsi}, {De}, {Dobie}, {Emery}, {Evans}, {Fox}, {Frail}, {Frohmaier}, {Goobar}, {Hallinan}, {Harrison}, {Helou}, {Hinderer}, {Ho}, {Horesh}, {Ip}, {Itoh}, {Kasen}, {Kim}, {Kuin}, {Kupfer}, {Lynch}, {Madsen}, {Mazzali}, {Miller}, {Mooley}, {Murphy}, {Ngeow}, {Nichols}, {Nissanke}, {Nugent}, {Ofek}, {Qi}, {Quimby}, {Rosswog}, {Rusu}, {Sadler}, {Schmidt}, {Sollerman}, {Steele}, {Williamson}, {Xu}, {Yan}, {Yatsu}, {Zhang}, \& {Zhao}}]{Kasliwal2017Science}
{Kasliwal}, M.~M., {Nakar}, E., {Singer}, L.~P., {et~al.} 2017, Science, 358, 1559, \dodoi{10.1126/science.aap9455}

\bibitem[{{Kuin} {et~al.}(2021){Kuin}, {Troja}, \& {Swift/UVOT Team}}]{Kuin2021GCN}
{Kuin}, N.~P.~M., {Troja}, E., \& {Swift/UVOT Team}. 2021, GRB Coordinates Network, 30572, 1

\bibitem[{{Lamb} {et~al.}(2022){Lamb}, {Nativi}, {Rosswog}, {Kann}, {Levan}, {Lundman}, \& {Tanvir}}]{Lamb2022Univ}
{Lamb}, G.~P., {Nativi}, L., {Rosswog}, S., {et~al.} 2022, Universe, 8, 612, \dodoi{10.3390/universe8120612}

\bibitem[{{Lazzati} {et~al.}(2018){Lazzati}, {Perna}, {Morsony}, {Lopez-Camara}, {Cantiello}, {Ciolfi}, {Giacomazzo}, \& {Workman}}]{Lazzati2018PhRvL}
{Lazzati}, D., {Perna}, R., {Morsony}, B.~J., {et~al.} 2018, \prl, 120, 241103, \dodoi{10.1103/PhysRevLett.120.241103}

\bibitem[{{Lazzati} {et~al.}(2003){Lazzati}, {Covino}, {di Serego Alighieri}, {Ghisellini}, {Vernet}, {Le Floc'h}, {Fugazza}, {Di Tomaso}, {Malesani}, {Masetti}, {Pian}, {Oliva}, \& {Stella}}]{Lazzati2003A&A}
{Lazzati}, D., {Covino}, S., {di Serego Alighieri}, S., {et~al.} 2003, \aap, 410, 823, \dodoi{10.1051/0004-6361:20031321}

\bibitem[{Li {et~al.}(2023)Li, Gao, Ai, \& Lei}]{Li2023MNRAS}
Li, J.-D., Gao, H., Ai, S., \& Lei, W.-H. 2023, Monthly Notices of the Royal Astronomical Society, stad2606, \dodoi{10.1093/mnras/stad2606}

\bibitem[{Li {et~al.}(2024)Li, Gao, Ai, \& Lei}]{Li2024submit}
---. 2024, submitted

\bibitem[{{Lyman} {et~al.}(2018){Lyman}, {Lamb}, {Levan}, {Mandel}, {Tanvir}, {Kobayashi}, {Gompertz}, {Hjorth}, {Fruchter}, {Kangas}, {Steeghs}, {Steele}, {Cano}, {Copperwheat}, {Evans}, {Fynbo}, {Gall}, {Im}, {Izzo}, {Jakobsson}, {Milvang-Jensen}, {O'Brien}, {Osborne}, {Palazzi}, {Perley}, {Pian}, {Rosswog}, {Rowlinson}, {Schulze}, {Stanway}, {Sutton}, {Th{\"o}ne}, {de Ugarte Postigo}, {Watson}, {Wiersema}, \& {Wijers}}]{Lyman2018NatAs}
{Lyman}, J.~D., {Lamb}, G.~P., {Levan}, A.~J., {et~al.} 2018, Nature Astronomy, 2, 751, \dodoi{10.1038/s41550-018-0511-3}

\bibitem[{{Mangano} {et~al.}(2007){Mangano}, {Holland}, {Malesani}, {Troja}, {Chincarini}, {Zhang}, {La Parola}, {Brown}, {Burrows}, {Campana}, {Capalbi}, {Cusumano}, {Della Valle}, {Gehrels}, {Giommi}, {Grupe}, {Guidorzi}, {Mineo}, {Moretti}, {Osborne}, {Pandey}, {Perri}, {Romano}, {Roming}, \& {Tagliaferri}}]{Mangano2007A&A}
{Mangano}, V., {Holland}, S.~T., {Malesani}, D., {et~al.} 2007, \aap, 470, 105, \dodoi{10.1051/0004-6361:20077232}

\bibitem[{{M{\'e}sz{\'a}ros} \& {Rees}(1997)}]{Meszaros1997ApJ}
{M{\'e}sz{\'a}ros}, P., \& {Rees}, M.~J. 1997, \apj, 476, 232, \dodoi{10.1086/303625}

\bibitem[{{Nakar} \& {Oren}(2004)}]{Nakar2004ApJ}
{Nakar}, E., \& {Oren}, Y. 2004, \apjl, 602, L97, \dodoi{10.1086/382729}

\bibitem[{{Narayan} \& {Kumar}(2009)}]{Narayan2009MNRAS}
{Narayan}, R., \& {Kumar}, P. 2009, \mnras, 394, L117, \dodoi{10.1111/j.1745-3933.2009.00624.x}

\bibitem[{{Nicuesa Guelbenzu} {et~al.}(2021{\natexlab{a}}){Nicuesa Guelbenzu}, {Klose}, {Schmidl}, \& {Rau}}]{Nicuesa2021GCN}
{Nicuesa Guelbenzu}, A., {Klose}, S., {Schmidl}, S., \& {Rau}, A. 2021{\natexlab{a}}, GRB Coordinates Network, 30574, 1

\bibitem[{{Nicuesa Guelbenzu} {et~al.}(2021{\natexlab{b}}){Nicuesa Guelbenzu}, {Klose}, {Schmidl}, {Rau}, \& {Kann}}]{Nicuesa2021GCNb}
{Nicuesa Guelbenzu}, A., {Klose}, S., {Schmidl}, S., {Rau}, A., \& {Kann}, D.~A. 2021{\natexlab{b}}, GRB Coordinates Network, 30584, 1

\bibitem[{{O'Connor} {et~al.}(2023){O'Connor}, {Troja}, {Ryan}, {Beniamini}, {van Eerten}, {Granot}, {Dichiara}, {Ricci}, {Lipunov}, {Gillanders}, {Gill}, {Moss}, {Anand}, {Andreoni}, {Becerra}, {Buckley}, {Butler}, {Cenko}, {Chasovnikov}, {Durbak}, {Francile}, {Hammerstein}, {van der Horst}, {Kasliwal}, {Kouveliotou}, {Kutyrev}, {Lee}, {Srinivasaragavan}, {Topolev}, {Watson}, {Yang}, \& {Zhirkov}}]{OConnorSci}
{O'Connor}, B., {Troja}, E., {Ryan}, G., {et~al.} 2023, Science Advances, 9, eadi1405, \dodoi{10.1126/sciadv.adi1405}

\bibitem[{{Peng} {et~al.}(2005){Peng}, {K{\"o}nigl}, \& {Granot}}]{Peng2005ApJ}
{Peng}, F., {K{\"o}nigl}, A., \& {Granot}, J. 2005, \apj, 626, 966, \dodoi{10.1086/430045}

\bibitem[{{Perley} {et~al.}(2011){Perley}, {Chandler}, {Butler}, \& {Wrobel}}]{Perley2011ApJ}
{Perley}, R.~A., {Chandler}, C.~J., {Butler}, B.~J., \& {Wrobel}, J.~M. 2011, \apjl, 739, L1, \dodoi{10.1088/2041-8205/739/1/L1}

\bibitem[{{Piro} \& {Kollmeier}(2018)}]{Piro2018ApJ}
{Piro}, A.~L., \& {Kollmeier}, J.~A. 2018, \apj, 855, 103, \dodoi{10.3847/1538-4357/aaaab3}

\bibitem[{{Roming} {et~al.}(2005){Roming}, {Kennedy}, {Mason}, {Nousek}, {Ahr}, {Bingham}, {Broos}, {Carter}, {Hancock}, {Huckle}, {Hunsberger}, {Kawakami}, {Killough}, {Koch}, {McLelland}, {Smith}, {Smith}, {Soto}, {Boyd}, {Breeveld}, {Holland}, {Ivanushkina}, {Pryzby}, {Still}, \& {Stock}}]{Roming2005SSRv}
{Roming}, P. W.~A., {Kennedy}, T.~E., {Mason}, K.~O., {et~al.} 2005, \ssr, 120, 95, \dodoi{10.1007/s11214-005-5095-4}

\bibitem[{{Rossi} {et~al.}(2004){Rossi}, {Lazzati}, {Salmonson}, \& {Ghisellini}}]{Rossi2004MNRAS}
{Rossi}, E.~M., {Lazzati}, D., {Salmonson}, J.~D., \& {Ghisellini}, G. 2004, \mnras, 354, 86, \dodoi{10.1111/j.1365-2966.2004.08165.x}

\bibitem[{{Sari} {et~al.}(1998){Sari}, {Piran}, \& {Narayan}}]{Sari1998ApJL}
{Sari}, R., {Piran}, T., \& {Narayan}, R. 1998, \apjl, 497, L17, \dodoi{10.1086/311269}

\bibitem[{{Stamatikos} {et~al.}(2021){Stamatikos}, {Barthelmy}, {Krimm}, {Laha}, {Lien}, {Markwardt}, {Palmer}, {Sakamoto}, {Troja}, \& {Ukwatta}}]{Stamatikos2021GCN}
{Stamatikos}, M., {Barthelmy}, S.~D., {Krimm}, H.~A., {et~al.} 2021, GRB Coordinates Network, 30580, 1

\bibitem[{{Swenson} {et~al.}(2013){Swenson}, {Roming}, {De Pasquale}, \& {Oates}}]{Swenson2013ApJ}
{Swenson}, C.~A., {Roming}, P.~W.~A., {De Pasquale}, M., \& {Oates}, S.~R. 2013, \apj, 774, 2, \dodoi{10.1088/0004-637X/774/1/2}

\bibitem[{{Troja} {et~al.}(2021){Troja}, {Ambrosi}, {D'Elia}, {Lien}, {Marshall}, {Sbarufatti}, {Tohuvavohu}, \& {Neil Gehrels Swift Observatory Team}}]{Troja2021GCN}
{Troja}, E., {Ambrosi}, E., {D'Elia}, V., {et~al.} 2021, GRB Coordinates Network, 30568, 1

\bibitem[{{Troja} {et~al.}(2018){Troja}, {Piro}, {Ryan}, {van Eerten}, {Ricci}, {Wieringa}, {Lotti}, {Sakamoto}, \& {Cenko}}]{Troja2018MNRAS}
{Troja}, E., {Piro}, L., {Ryan}, G., {et~al.} 2018, \mnras, 478, L18, \dodoi{10.1093/mnrasl/sly061}

\bibitem[{{van Paradijs} {et~al.}(1997){van Paradijs}, {Groot}, {Galama}, {Kouveliotou}, {Strom}, {Telting}, {Rutten}, {Fishman}, {Meegan}, {Pettini}, {Tanvir}, {Bloom}, {Pedersen}, {N{\o}rdgaard-Nielsen}, {Linden-V{\o}rnle}, {Melnick}, {Van der Steene}, {Bremer}, {Naber}, {Heise}, {in't Zand}, {Costa}, {Feroci}, {Piro}, {Frontera}, {Zavattini}, {Nicastro}, {Palazzi}, {Bennett}, {Hanlon}, \& {Parmar}}]{vanParadijs1997Natur}
{van Paradijs}, J., {Groot}, P.~J., {Galama}, T., {et~al.} 1997, \nat, 386, 686, \dodoi{10.1038/386686a0}

\bibitem[{{Wang} {et~al.}(2014){Wang}, {Lei}, {Wang}, {Zou}, {Zhang}, {Gao}, \& {Huang}}]{Wang2014}
{Wang}, J.-Z., {Lei}, W.-H., {Wang}, D.-X., {et~al.} 2014, \apj, 788, 32, \dodoi{10.1088/0004-637X/788/1/32}

\bibitem[{{Wang} {et~al.}(2015){Wang}, {Zhang}, {Liang}, {Gao}, {Li}, {Deng}, {Qin}, {Tang}, {Kann}, {Ryde}, \& {Kumar}}]{wang2015}
{Wang}, X.-G., {Zhang}, B., {Liang}, E.-W., {et~al.} 2015, \apjs, 219, 9, \dodoi{10.1088/0067-0049/219/1/9}

\bibitem[{{Wu} {et~al.}(2003){Wu}, {Dai}, {Huang}, \& {Lu}}]{Wu2003MNRAS}
{Wu}, X.~F., {Dai}, Z.~G., {Huang}, Y.~F., \& {Lu}, T. 2003, \mnras, 342, 1131, \dodoi{10.1046/j.1365-8711.2003.06602.x}

\bibitem[{{Wu} {et~al.}(2005){Wu}, {Dai}, {Huang}, \& {Lu}}]{Wu2005MNRAS}
---. 2005, \mnras, 357, 1197, \dodoi{10.1111/j.1365-2966.2005.08685.x}

\bibitem[{{Xiao} {et~al.}(2017){Xiao}, {Liu}, {Dai}, \& {Wu}}]{Xiao2017ApJL}
{Xiao}, D., {Liu}, L.-D., {Dai}, Z.-G., \& {Wu}, X.-F. 2017, \apjl, 850, L41, \dodoi{10.3847/2041-8213/aa9b2b}

\bibitem[{{Zhang} {et~al.}(2006){Zhang}, {Fan}, {Dyks}, {Kobayashi}, {M{\'e}sz{\'a}ros}, {Burrows}, {Nousek}, \& {Gehrels}}]{Zhang2006ApJ}
{Zhang}, B., {Fan}, Y.~Z., {Dyks}, J., {et~al.} 2006, \apj, 642, 354, \dodoi{10.1086/500723}

\bibitem[{{Zhang} {et~al.}(2018){Zhang}, {Zhang}, {Sun}, {Lei}, {Gao}, {Li}, {Shao}, {Zhao}, {Hu}, {L{\"u}}, {Wu}, {Fan}, {Wang}, {Castro-Tirado}, {Zhang}, {Yu}, {Cao}, \& {Liang}}]{Zhang2018NatureCommunications}
{Zhang}, B.~B., {Zhang}, B., {Sun}, H., {et~al.} 2018, Nature Communications, 9, 447, \dodoi{10.1038/s41467-018-02847-3}

\bibitem[{{Zheng} {et~al.}(2021){Zheng}, {Filippenko}, \& {KAIT GRB Team}}]{Zheng2021GCN}
{Zheng}, W., {Filippenko}, A.~V., \& {KAIT GRB Team}. 2021, GRB Coordinates Network, 30582, 1

\bibitem[{{Zhou} {et~al.}(2024){Zhou}, {Zhu}, {Lei}, {Fu}, {Xie}, \& {Xu}}]{Zhou2023}
{Zhou}, C., {Zhu}, Z.-P., {Lei}, W.-H., {et~al.} 2024, \apj, 963, 66, \dodoi{10.3847/1538-4357/ad20f3}

\end{thebibliography}
\bibliographystyle{aasjournal}



\end{document}